\documentclass{jfm}

\usepackage{graphicx}
\usepackage{newtxtext}
\usepackage{newtxmath}
\usepackage{natbib}
\usepackage{hyperref}
\hypersetup{
    colorlinks = true,
    urlcolor   = blue,
    citecolor  = black,
}

\newcommand{\RomanNumeralCaps}[1]
\linenumbers

\usepackage{bm,upgreek}
\usepackage{color,float,colortbl}
\usepackage[table,dvipsnames]{xcolor}

\definecolor{mypink1}{rgb}{0.858, 0.188, 0.478}
\definecolor{myred}{RGB}{196, 30, 58}
\definecolor{myred}{RGB}{210, 4, 45}
\definecolor{mypink3}{cmyk}{0, 0.7808, 0.4429, 0.1412}
\definecolor{mygray}{gray}{0.6}

\definecolor{Gray}{gray}{0.9}

\providecommand\bnabla{\mathbf{\nabla}}

\renewcommand{\d}{\mathrm{d}}

\newcommand{\vetu}[0]{\mbox{\bf{\em{u}}}}

\newcommand{\vetx}[0]{\mbox{\bf{\em{x}}}}
\newcommand{\vetz}[0]{\mbox{\bf{\em{z}}}}
\newcommand{\vetxh}[0]{\hat{\mbox{\bf{\em{x}}}}}

\newcommand{\vetk}[0]{\mbox{\bf{\em{k}}}}

\title{Stability of plane Couette and Poiseuille flows rotating about the streamwise axis}

\author{Geert Brethouwer
\corresp{\email{geert@kth.se}}}
\affiliation{Department of Engineering Mechanics, KTH
Royal Institute of Technology, SE-10044 Stockholm, Sweden}

\begin{document}
\maketitle

\begin{abstract}
We study the stability of plane Poiseuille flow (PPF) and plane Couette flow (PCF) subject to streamwise system rotation using linear stability analysis and direct numerical simulations. The linear stability analysis reveals two asymptotic regimes depending on the non-dimensional rotation rate ($Ro$): a low-$Ro$ and a high-$Ro$ regime. In the low-$Ro$ regime, the critical Reynolds number $Re_c$ and critical streamwise wavenumber $\alpha_c$ are proportional to $Ro$, while the critical spanwise wavenumber $\beta_c$ is constant. In the high-$Ro$ regime, as $Ro \rightarrow \infty$, we find $Re_c = 66.45$ and $\beta_c = 2.459$ for streamwise rotating PPF, and $Re_c = 20.66$ and $\beta_c = 1.558$ for streamwise rotating PCF, with $\alpha_c\propto 1/Ro$. Our results for streamwise rotating PPF match previous findings by \citet{Masuda}. Interestingly, the critical values of $\beta_c$ and $Re_c$ at $Ro \rightarrow \infty$ in \emph{streamwise} rotating PPF and PCF coincide with the minimum $Re_c$ reported by \citet{Lezius} and \citet{Wall} for \emph{spanwise} rotating PPF at $Ro=0.3366$ and PCF at $Ro=0.5$. We explain this similarity through an analysis of the perturbation equations. Consequently, the linear stability of streamwise rotating PCF at large $Ro$ is closely related to that of spanwise rotating PCF and Rayleigh–B{\'e}nard convection, with $Re_c = \sqrt{Ra_c}/2$, where $Ra_c$ is the critical Rayleigh number. To explore the potential for subcritical transitions, direct numerical simulations were performed. At low $Ro$, a subcritical transition regime emerges, characterized by large-scale turbulent-laminar patterns in streamwise rotating PPF and PCF. However, at higher $Ro$, subcritical transitions do not occur and the flow relaminarizes for $Re < Re_c$. Furthermore, we identify a narrow $Ro$-range where turbulent-laminar patterns develop under supercritical conditions.
\end{abstract}

\section{Introduction}

Wall-bounded shear flows such as
plane Poiseuille or channel flow (PPF) and plane Couette flow (PCF)
subject to system rotation display many interesting physical phenomena, for example,
turbulent-laminar patterns \citep{Brethouwer2012,Brethouwer2017}, recurring bursts of turbulence
\citep{Brethouwer2014,Brethouwer2016}, large-scale structures \citep{Gai,Brethouwer2017},
multiple states \citep{YangXia2021} and strong increases in momentum
and heat transfer \citep{Brauckmann2016,Brethouwer2021,Brethouwer2023}.
Studying the stability of such flows subject
to system rotation in various directions may help to
understand rotating shear flows in engineering applications.

The stability of PPF and PCF with and without 
spanwise system rotation has been studied extensively, 
see e.g. \cite{Schmid,Hart,Hung,Lezius,Wall,Daly,Nagata2021}.
From now on, we will abbreviate non-rotating PCF and PPF to NPCF and NPPF, respectively,
and PCF and PPF subject to spanwise system rotation to ZPCF and ZPPF, respectively.
NPCF is linearly stable at any Reynolds number $Re$, whereas in NPPF
two-dimensional Tollmien-Schlichting (TS) modes with
$\beta=0$ are linearly unstable for $Re \geq 5772.3$ \citep{Schmid}. Here, and in the following
$Re = U_{cl}\delta/\nu$ for PPF and $Re = U_w\delta/\nu$ for PCF, where
$U_{cl}$ is the centerline velocity, $U_w$ is the velocity of the two walls moving in
opposite directions, $\delta$ is the half gap-width and $\nu$ is the viscosity. 
Subscript $c$ is used to denote
values at critical condition for linear instabilities.
Further, $\alpha$ and $\beta$ are the streamwise and spanwise wavenumbers, 
respectively, nondimensionalized by $\delta$.

Spanwise rotation can drastically reduce
the critical Reynolds number $Re_c$ of PPF. 
\citet{Lezius} and \citet{Alfredsson} did
a linear stability analysis (LSA) of
ZPPF assuming two-dimensional perturbations with $\alpha=0$
and found that the minimum critical Reynolds number is $Re_c=66.40$ at $Ro_c=1/3$. 
Here, and in the following $Ro=2\Omega\delta/U_{cl}$ for PPF
and $Ro=2\Omega\delta/U_w$ for PCF, where $\Omega$ is the imposed system rotation rate.
\citet{Wall} extended the LSA to three-dimensional perturbations and confirmed
that at low $Re$, ZPPF is most unstable to perturbations
with $\alpha=0$. They recomputed the critical values and
found the lowest $Re_c = 66.448$
at $Ro=0.3366$ with $\beta_c=2.459$.

\citet{Lezius} also pointed out the similarity between the linear perturbation
equations of ZPCF and
Rayleigh-Benard convection between two flat plates.
From that similarity follows $16Re^2_c Ro(1-Ro)=Ra_c$ and
$\beta_c=3.117/2=1.558$ when $Ro>0$, 
where $Ra_c=1707.762$ is the critical Rayleigh number
\citep{Chandrasekhar} and $Ro>0$ corresponds to anticyclonic rotation.
The factors 16 and 2 in the relations for $Re_c$ and $\beta_c$
arise when the half gap width $\delta$ is used for nondimensionalization
instead of the gap width. The previous relation shows that
the minimum $Re_c=20.6625$ of ZPCF occurs at $Ro=0.5$.
The non-normality
of the linearized Navier-Stokes operator of PPF and PCF
can explain the strong reduction of $Re_c$ 
by spanwise rotation \citep{Jose2020}.

Experiments \citep{Alfredsson,Tsukahara} show that
streamwise vortices develop in ZPPF and ZPCF
slightly above $Re_c$.
The vortices are steady
and turbulent motions are absent at these low $Re$,
but the vortices become three-dimensional and unstable when $Re$ increases
\citep{Yang,Finlay,Nagata,Tsukahara,Daly,Nagata2021}, and
turbulence sets in at sufficiently high $Re$ 
\citep{Tsukahara,Salewski,Jose2017,Brethouwer2017,Brethouwer2021}.

The effect of system rotation about axes other than the spanwise axis on PPF
has also been investigated.
\citet{Wu} studied the effect of system rotation
with various rotation axis directions
on turbulent PPF using direct
numerical simulation (DNS).
PPF subject to streamwise system rotation,
abbreviated as XPPF, has also been investigated. 
DNS \citep{Oberlack,Yang2010,Yang2018,Yang2020,Yu,Hu2023,Hu2024} 
and experiments \citep{Recktenwald} of turbulent XPPF
show a secondary mean flow and
distinct Taylor-G{\"o}rtler vortices,
which are inclined to the streamwise direction.

Instabilities in XPPF have been studied via
a LSA and non-linear analysis by \citet{Masuda}.
They used a nondimensional rotation rate
$\Upomega^* = 2\Omega \delta^2/\nu$, which can be related
to $Ro$ by noting that $\Upomega^*=Re\,Ro$.
\citet{Masuda} observed two asymptotic neutral stability regimes
for three-dimensional perturbations in the LSA;
one at low $Ro$ with $Re_c = 33.923/Ro$,
and one at high $Ro$ with $Re_c = 66.45$
and $\beta_c\simeq 2.5$ and $\alpha_c$ decreasing with $Ro$.
At very low $Ro$, XPPF is most unstable to two-dimensional
TS modes.
The values of $Re_c$ and $\beta_c$ in
XPPF at high $Ro$ are remarkably similar to the minimum
$Re_c = 66.448$ and $\beta_c = 2.4592$
occurring at $Ro=0.3366$ in ZPPF \citep{Wall}.
\citet{Masuda} did not comment on this similarity,
but we will show that it is not a coincidence.

LSA does not always accurately predict a critical $Re$ for transition. 
Disturbances can exhibit 
transient energy growth in linearly stable flows
due to the non-normality of the linearized Navier-Stokes operator \citep{Grossmann}, 
potentially triggering a subcritical transition \citep{Orszag,Daviaud}. 
Consequently, the energy method has been employed to determine an
energy-based Reynolds number threshold $Re_E$,
below which all disturbances monotonically decay \citep{Boeck}.
This approach has been applied to NPPF and NPCF
\citep{Orr,Joseph1976,Busse1969,Busse1972,Falsaperla}, showing that
$Re_E$ in NPPF is two orders of magnitude lower than $Re_c$.
However, even if transient growth occurs, a flow may relaminarize if disturbances do not grow sufficiently to trigger a subcritical transition. 
Hence, $Re_E$ can be significantly lower than the critical $Re$ 
below which a shear flow remains laminar \citep{Fuentes}.

Although $Re_E$ is a conservative measure,
it is observed that turbulence can persist
in NPPF at $Re$ much lower than $Re_c$ and in NPCF at finite $Re$.
However, below some $Re$ threshold, NPPF and NPCF
are not uniformly turbulent but transitional.
Intermittent turbulence, sometimes
forming large-scale oblique bands with alternating
laminar-like and turbulent-like flow, can develop in a range of $Re$
\citep{Tuckerman,Duguet,Shimuzu}. 
The flows eventually become laminar at lower $Re$, 
regardless of the initial conditions, and
fully turbulent at higher $Re$.

Subcritical transition has also
been studied in rotating shear flows. 
\citet{Jose2017}
investigated transient growth in ZPPF and showed that
the critical Reynolds number for such growth is almost independent of $Ro$,
decreasing from 51 at low rotation to 41 at high rotation.
These values are far below $Re_c$, 
both for $Ro\lesssim 10^{-2}$ and for large $Ro$.
Their DNS confirmed that subcritical transition can occur
at low $Ro$. DNS and experiments of ZPCF with cyclonic
rotation also show subcritical transition
and turbulent-laminar patterns in
some $(Re,Ro)$ range \citep{Tsukahara,Brethouwer2012}.

Subcritical transition to turbulence in XPPF has not, to our knowledge,
been examined in detail. The values of $Ro$ considered in DNS
of XPPF are $Ro\approx 0.1$ \citep{Oberlack} and higher, where $Re_c$ 
is much lower than the $Re$ in the DNS, so the transition is supercritical.
The energy and linear stability of spiral flow between concentric cylinders,
rotating and sliding relative to one another, with or without a uniform
axial pressure gradient, 
have been investigated by \cite{Joseph1970,Hung,Joseph1976}.
XPCF and XPPF represent two limiting narrow-gap cases of spiral flow with equal
rotation rates of the inner and outer cylinders.

In summary, the stability of ZPPF and ZPCF
has been extensively studied,
whereas the stability of XPPF and XPCF has received much less attention. 
Only \citet{Masuda}
and briefly \citet{Joseph1970,Hung,Joseph1976} have studied these two flow cases.
Many aspects of the behaviour of the critical modes as
well as the potential of subcritical transition
remain unclear.
In this study, we investigate the stability of XPPF
and XPCF with the aim of obtaining a deeper physical understanding of these
flows.
We perform both LSA and DNS to examine the possibility of
subcritical transition.

\section{Methodology: linear stability analysis and direct numerical simulations}

\subsection{Configuration and governing equations}

\begin{figure}
\begin{tabular}{cc}
\includegraphics[width=6.5cm]{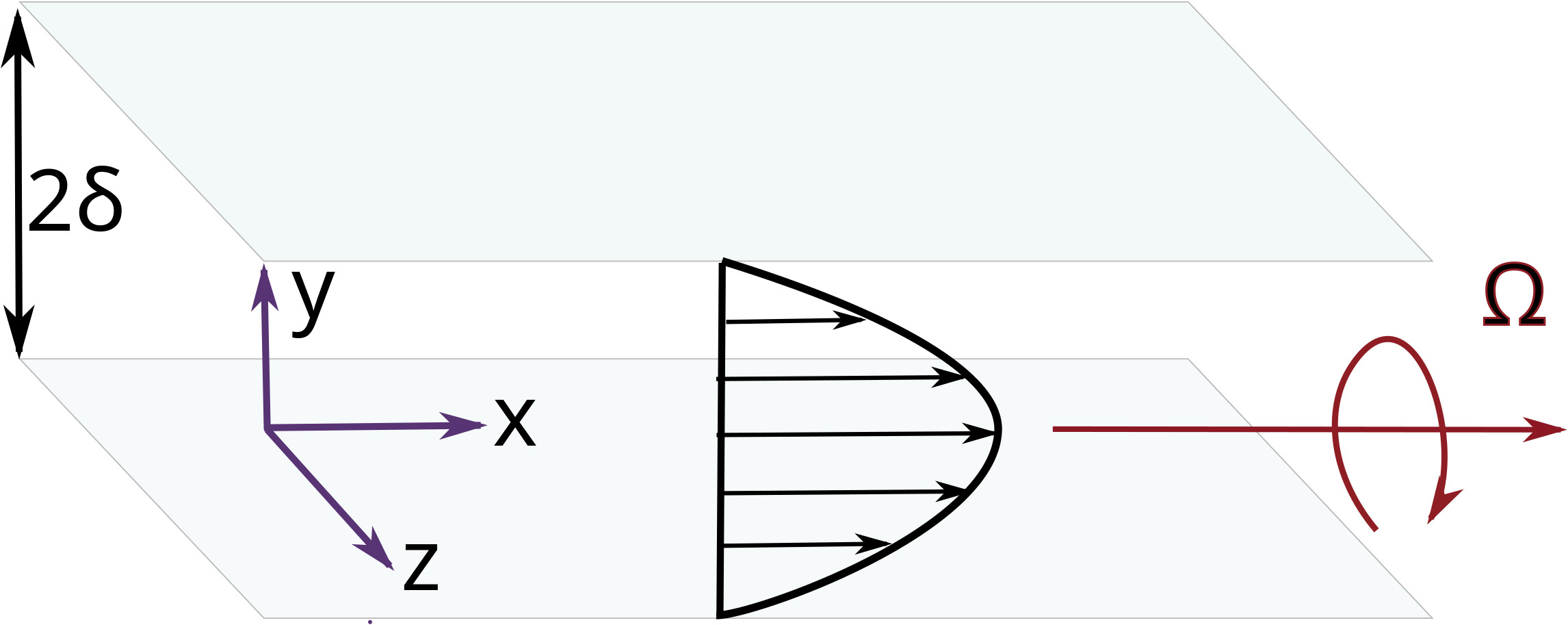}
&
\includegraphics[width=6.5cm]{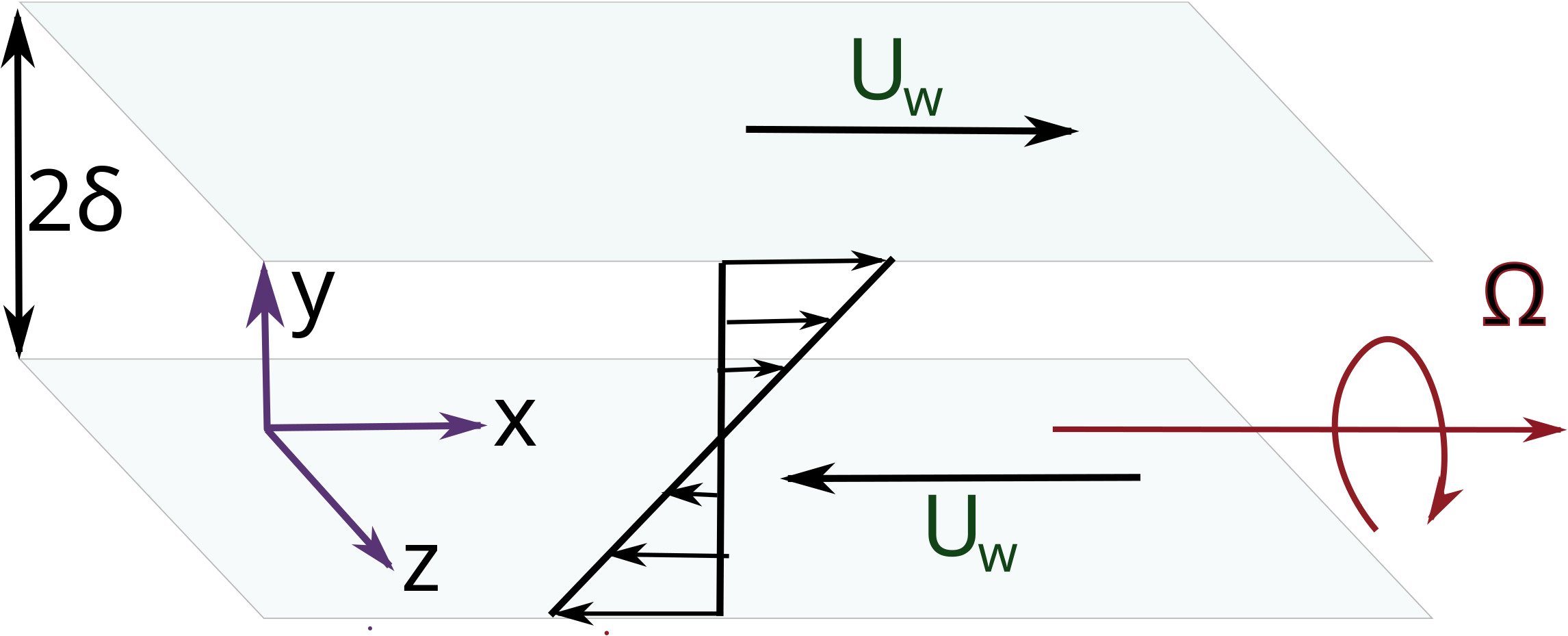}\\
(a) & (b)
\end{tabular}
\caption{(a) XPPF and (b) XPCF configurations.\label{config}}
\end{figure}

We investigate the stability of viscous incompressible
PCF and pressure driven PPF
subject to constant system rotation about the streamwise axis, i.e.,
XPCF and XPPF,
as illustrated in figure \ref{config}.
The streamwise, wall-normal and spanwise coordinates nondimensionalized
by the half gap-width $\delta$ are denoted by $x$, $y$ and $z$, respectively,
and the corresponding velocity components by $u$, $v$ and $w$, respectively.
The two infinite plane no-slip walls are at $y=\pm 1$.
The velocity $\vetu$ in the rotating frame of reference
in both flow cases is governed by the nondimensional Navier-Stokes equations
\begin{equation}
\frac{\p \vetu}{\p t} + \vetu \cdot \bnabla \vetu = - \bnabla p + \frac{1}{Re} \bnabla^2 \vetu - Ro (\vetxh \times \vetu),~~~
\bnabla \cdot \vetu = 0,
\label{gov_eq}
\end{equation}
where $\vetxh$ is the unit vector in the $x$-direction.
The last term in the momentum equation is the Coriolis force and
the centrifugal force is absorbed in a modified pressure $p$.
The laminar streamwise velocity profile in XPPF, given by $U=1-y^2$,
and in XPCF, given by $U=y$, is not affected by rotation. 

\subsection{Linear stability analysis}

We use standard linear stability methodology and linearize the
governing equations (\ref{gov_eq}).
Introducing wall-normal velocity
$v(\vetx,t)$ and wall-normal vorticity $\eta(\vetx,t)$ perturbations gives
\begin{subequations}
\begin{align}
\left [ \left ( \frac{\p}{\p t} + U \frac{\p}{\p x} \right ) \nabla^2 - U'' \frac{\p}{\p x}
- \frac{1}{Re} \nabla^4 \right ] v + Ro \frac{\p \eta}{\p x} &= 0 \label{lin_eqa}\\
\left [ \left ( \frac{\p}{\p t} + U \frac{\p}{\p x} \right ) 
- \frac{1}{Re} \nabla^2 \right ] \eta +
\left [ U'\frac{\p}{\p z}- Ro \frac{\p}{\p x} \right ] v &= 0,
\label{lin_eqb}
\end{align}
\end{subequations}
where $U' = \d U/\d y$ and $U'' = \d^2 U /\d y^2$ and boundary conditions
$v = \p v/\p y = \eta = 0$ at the walls.
Assuming wave-like perturbations 
$v(\vetx,t) = \hat{v}(y) e^{i(\alpha x + \beta z - \omega t)}$ and
$\eta(\vetx,t) = \hat{\eta}(y) e^{i(\alpha x + \beta z - \omega t)}$
with wavenumber vector $\vetk=(\alpha,\beta)$
leads to the following eigenvalue problem in matrix form
\begin{equation}
-i\omega\begin{pmatrix}
D^2-k^2 & 0 \\
0 & 1
\end{pmatrix}
\begin{pmatrix}
\hat{v} \\
\hat{\eta}
\end{pmatrix}
+
\begin{pmatrix}
\mathcal{L}_{OS} & \mathcal{L}_R \\
\mathcal{L}_C & \mathcal{L}_{SQ}
\end{pmatrix}
\begin{pmatrix}
\hat{v} \\
\hat{\eta}
\end{pmatrix}
=
\begin{pmatrix}
0 \\
0
\end{pmatrix},
\label{eigen}
\end{equation}
where the Orr-Sommerfeld and Squire operators $\mathcal{L}_{OS}$ and $\mathcal{L}_{SQ}$
and operators $\mathcal{L}_R$ and $\mathcal{L}_C$ are given by
\begin{subequations}
\begin{align}
\mathcal{L}_{OS} &= i \alpha U (D^2-k^2) - i \alpha U'' - \frac{1}{Re} (D^2-k^2)^2 \\
\mathcal{L}_R &= i\alpha Ro \\
\mathcal{L}_C & = i (\beta U' - \alpha Ro) \\
\mathcal{L}_{SQ} &= i \alpha U - \frac{1}{Re} (D^2-k^2),
\end{align}
\label{operators}
\end{subequations}
with $k^2 = \alpha^2 + \beta^2$ and $D(.) = \d(.)/\d y$ and boundary conditions
$\hat{v}(y) = D \hat{v}(y)= \hat{\eta}(y) =0$ at $y=\pm 1$.
This eigenvalue problem (\ref{eigen}) for $\omega$ with eigenvalues 
$\hat{v}(y)$ and $\hat{\eta}(y)$ for XPPF and XPCF
is discretized using Chebyshev polynomials 
and solved with \textsc{Matlab}{} routines. 
The imaginary part $\omega_i$ of the complex
eigenvalue $\omega$ gives the nondimensional growth rate of the perturbations.
Convergence has been checked by changing
the number of collocation points.

\subsection{Direct numerical simulations}

We also carry out DNS to investigate the stability of XPPF and XPCF,
using a pseudospectral code that solves equations (\ref{gov_eq})
with Fourier expansions and periodic boundary conditions
in $x$- and $z$- and Chebyshev polynomials in $y$-direction and no-slip conditions
at the walls \citep{Chevalier}. In the DNS of XPPF the flow rate is
fixed. The code has been used in many previous studies, e.g.
\citep{Brethouwer2012,Brethouwer2017,Brethouwer2021}.

\subsection{Validation of the linear stability analysis}

LSA results for XPPF agree with those of \citet{Masuda}, as we will show later.
To further validate our LSA we have carried out
DNS of XPCF and XPPF with small
initial perturbations at three $Ro$
and $Re$ slightly above $Re_c$. We compared the growth rate of
the velocity fluctuations with that
of the most unstable mode predicted by LSA. The growth rates
match, as shown in Appendix \ref{Ap_A}.

\section{Results: linear stability analysis}

\begin{figure}
\begin{center}
\includegraphics[width=10.0cm]{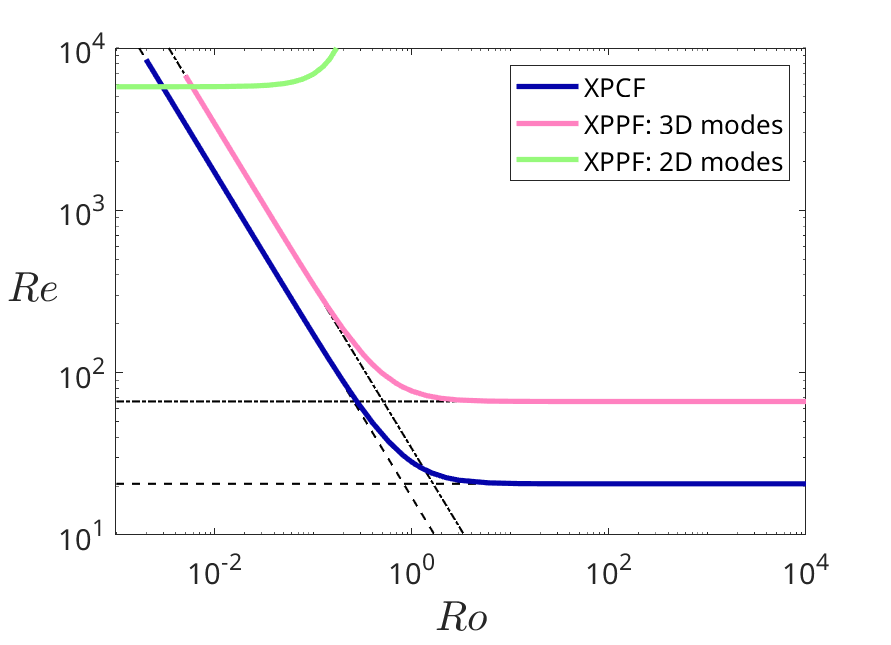}
\end{center}
\caption{Neutral stability curves of three-dimensional modes in XPCF
and the two-dimensional $\beta=0$ and three-dimensional modes in XPPF.
Horizontal dashed line, $Re=\sqrt{1707.762}/2$; dash-dotted line, $Re=66.45$.
Sloped dashed line, $Re=17/Ro$; dash-dotted line, $Re=33.923/Ro$.
\label{neutral}}
\end{figure}

We first discuss the LSA results.
Figure \ref{neutral} shows the neutral stability curves of XPPF and XPCF.
The most unstable mode in XPCF is three-dimensional
due to system rotation and the same applies to XPPF, 
except at very low $Ro$ when a two-dimensional TS mode
with $\beta=0$ is most unstable. 
The neutral stability curve of this TS mode, which
converges for $Ro\rightarrow 0$ to the critical
Reynolds number $Re_c=5772.2$ of NPPF, is also shown. 
According to figure \ref{neutral} we can distinguish two
asymptotic neutral stability regimes for three-dimensional perturbations
in XPCF and XPPF; 
a low-$Ro$ regime with $Re_c \propto 1/Ro$ at $Ro\rightarrow 0$,
and a high-$Ro$ regime with
$Re_c$ approaching a low constant value at $Ro\rightarrow \infty$.
The transition between these two regimes is at $Ro \sim O(1)$.
Since $Ro$ expresses the ratio of system rotation
to mean shear rotation, we can call
the regime with $Ro\gg 1$ a rotation dominated regime
and the regime with $Ro\ll 1$ a shear dominated regime.

\citet{Masuda} already identified 
these two asymptotic regimes for XPPF.
They found
$Re_c = 33.923/Ro$ in the low-$Ro$ regime,
and $Re_c = 66.45$ in the high-$Ro$ regime,
shown by dash-dotted
lines in figure \ref{neutral}, which match our LSA results.
In XPCF, $Re_c \simeq 17/Ro$ when $Ro\rightarrow 0$
and $Re_c = 20.6625$ when $Ro\rightarrow\infty$, 
shown by dashed lines in \ref{neutral}. The latter $Re_c$ is equal to
the minimum critical Reynolds number
$Re_c = \sqrt{1707.762}/2 = 20.6625$ in ZPCF occurring at $Ro=1/2$,
which is explained in the next section. 
In the high-$Ro$ regime of XPPF something similar happens
since $Re_c$ approaches $66.45$, which is equal
to the minimum $Re_c$ in ZPPF occurring at $Ro=0.3366$ \citep{Wall}.

\begin{figure}
\begin{center}
\begin{tabular}{cc}
\includegraphics[width=6.5cm]{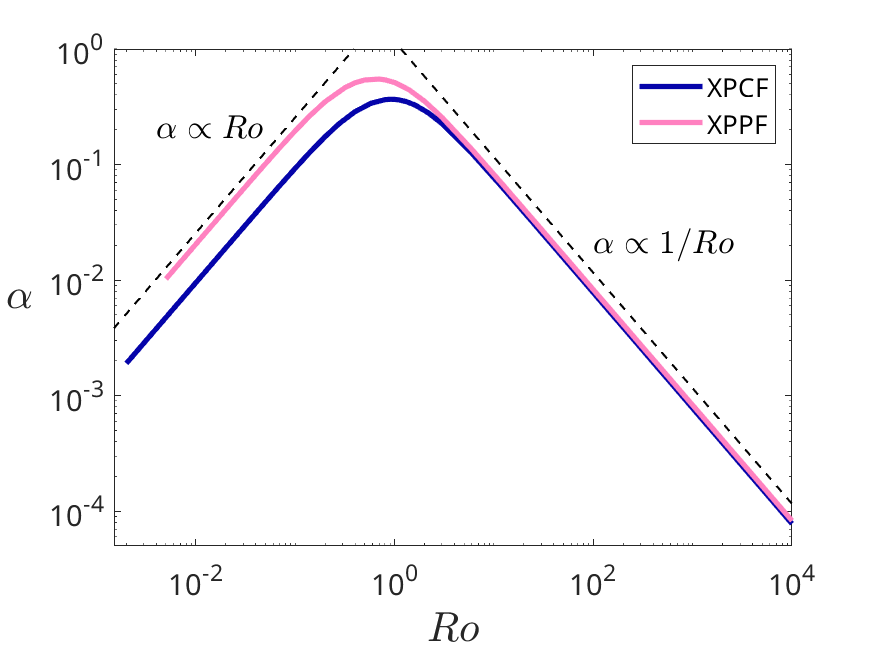}
&
\includegraphics[width=6.5cm]{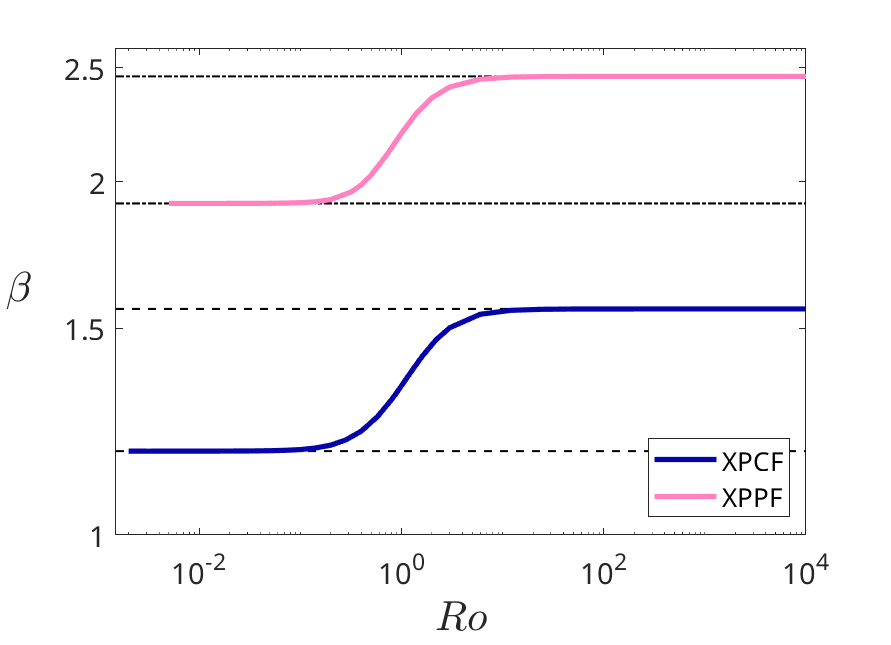}\\
(a) & (b)
\end{tabular}
\includegraphics[width=6.5cm]{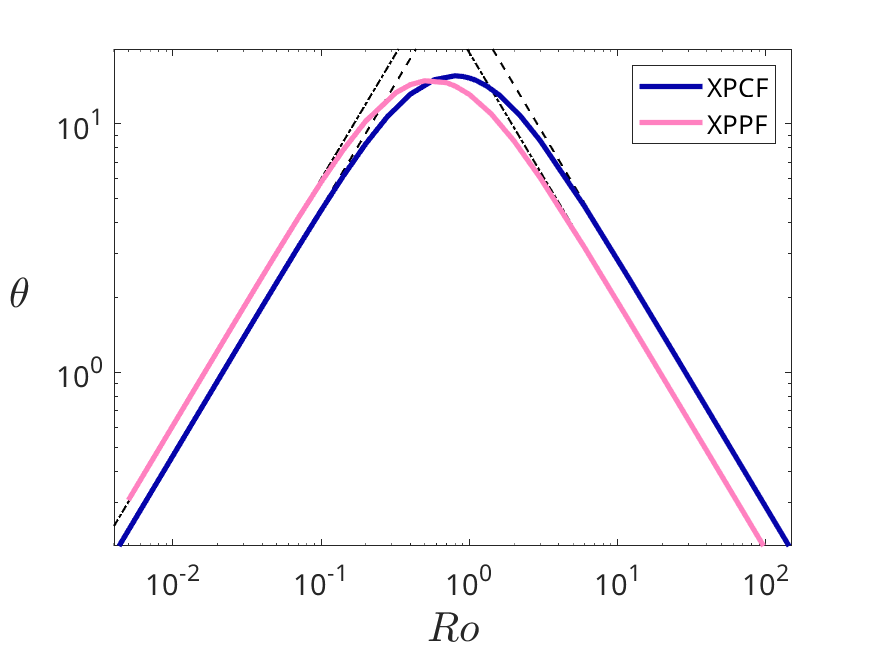}\\
(c)
\end{center}
\caption{The critical wavenumbers (a) $\alpha_c$ and (b) $\beta_c$ 
and (c) angle $\theta$ of the wavenumber vector $\vetk_c=(\alpha_c,\beta_c)$ 
with the $z$-axis as a function of $Ro$ in XPPF and XPCF.
In (b) dashed lines, $\beta=1.179$ and $\beta=1.558$; dash-dotted lines,
$\beta=1.917$ and $\beta=2.459$.
In (c) dashed lines, $\theta=0.5/Ro$ and $\theta=0.8Ro$; dash-dotted lines,
$\theta=0.3366/Ro$ and $\theta=1.05Ro$.
\label{angle}}
\end{figure}

Figure \ref{angle}(a,b) shows the wavenumbers of the critical
three-dimensional mode $\alpha_c$ and $\beta_c$
at neutral stability conditions, and 
figure \ref{angle}(c) the angle $\theta = \arctan(\alpha_c/\beta_c)$ of the wavenumber vector
$\vetk_c=(\alpha_c,\beta_c)$ with the $z$-axis as
a function of $Ro$ in XPPF and XPCF.
The critical spanwise wavenumber
assumes a constant but different value in the low-$Ro$ 
and high-$Ro$ regimes, and changes at $Ro\sim O(1)$.
In XPPF, $\beta_c = 2.459$ when $Ro\rightarrow \infty$,
which is the same $\beta_c$ as in ZPPF
at the minimum $Re_c$ at $Ro=0.3366$ \citep{Wall}. 
Similarly, in XPCF, $\beta_c = 1.558$ when $Ro\rightarrow \infty$,
which is the same $\beta_c$ as in ZPCF at the minimum $Re_c$ at $Ro=0.5$
\citep{Lezius}, which in turn is the same critical wavenumber as
in Rayleigh-B{\'e}nard convection \citep{Chandrasekhar}. 
The angle $\theta_c$ and $\alpha_c$ assume a maximum value at
$Ro\sim O(1)$ and decrease as $\theta_c , \alpha_c \propto 1/Ro$
as $Ro\rightarrow \infty$, and increase as $\theta_c , \alpha_c \propto Ro$
at $Ro\rightarrow 0$ in XPPF and XPCF.
The critical vortical structures
have thus the largest inclination angle with respect to the streamwise direction
when $Ro \sim O(1)$ and system rotation and
mean shear rotation are of the same order, and
become more aligned with the streamwise direction
when $Ro \rightarrow 0$ and $Ro\rightarrow \infty$.
The observed alignment of the vortices 
with the $x$-axis for $Ro\rightarrow \infty$ 
conforms to the Taylor-Proudman theorem.

\begin{figure}
\begin{center}
\begin{tabular}{cc}
\includegraphics[width=6.5cm]{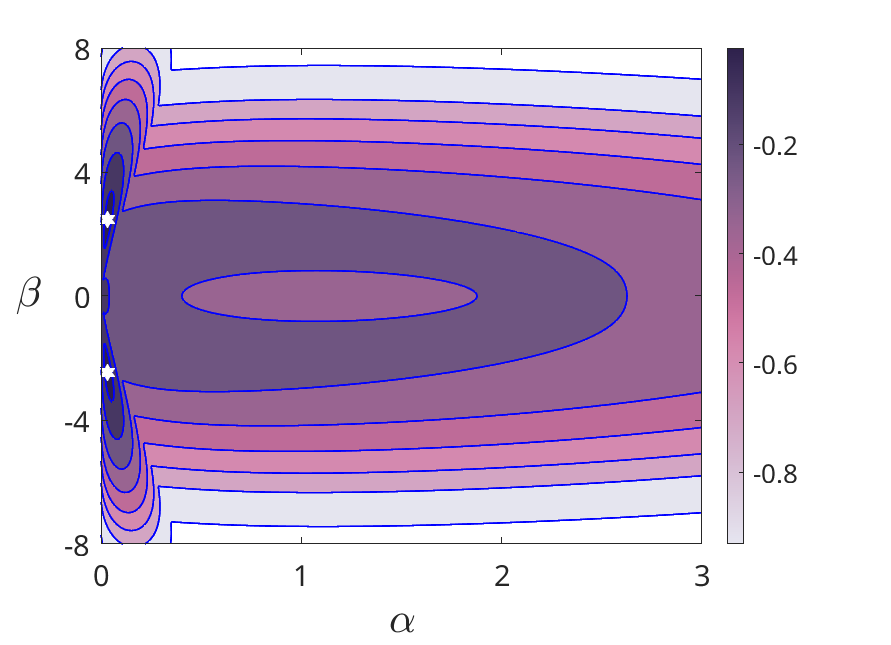} &
\includegraphics[width=6.5cm]{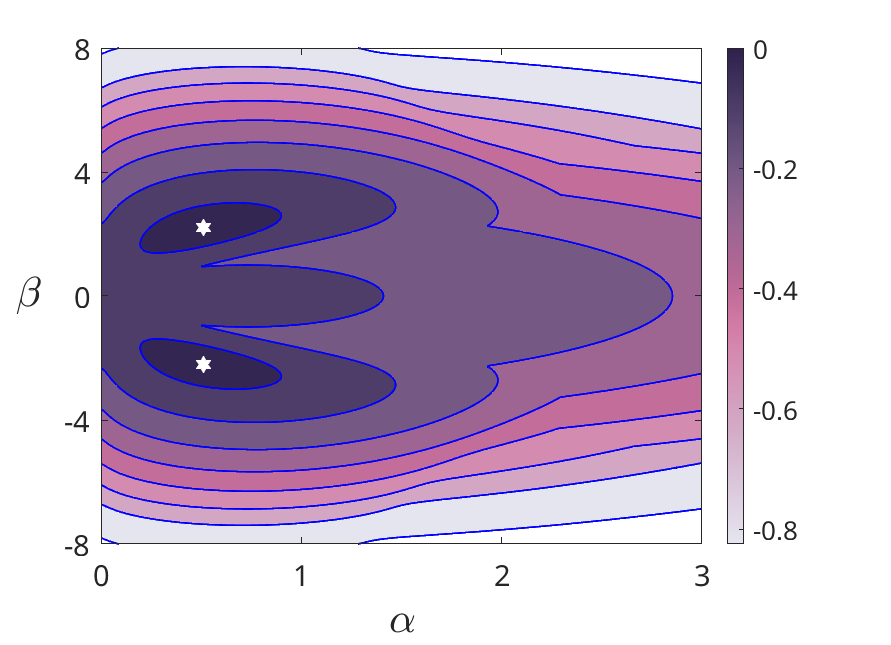}\\
(a) & (b)\\
\includegraphics[width=6.5cm]{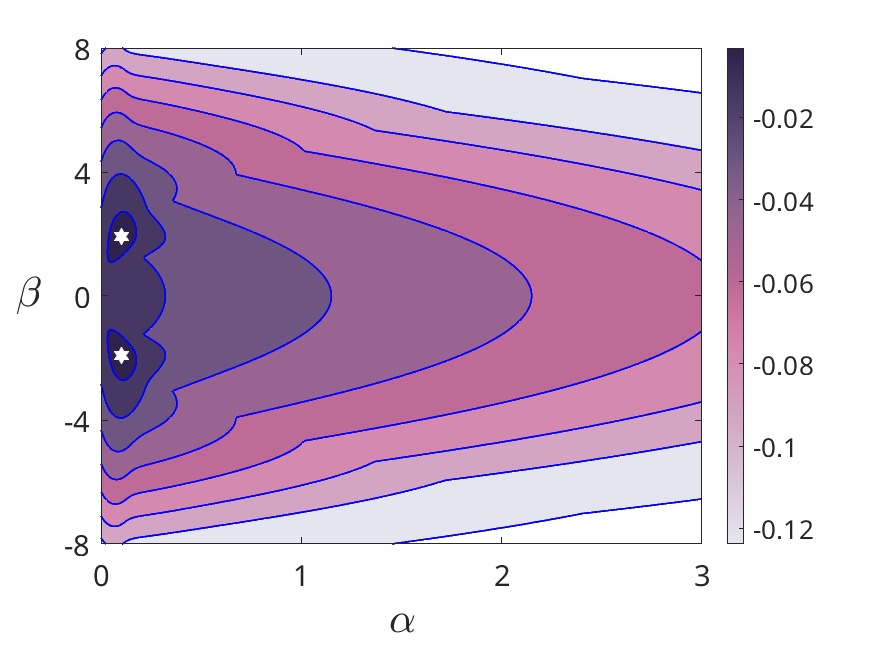} &
\includegraphics[width=6.5cm]{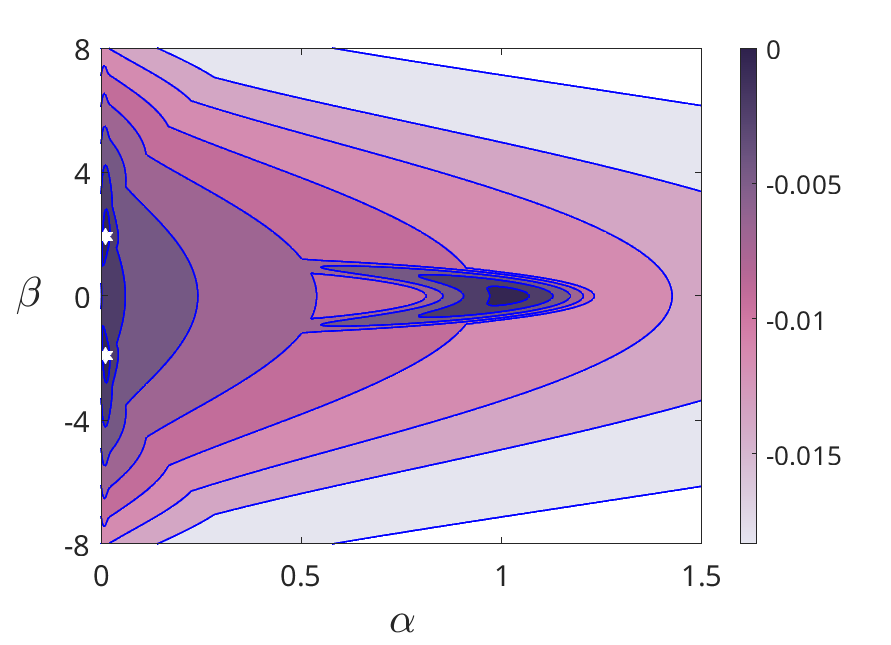}\\
(c) & (d)
\end{tabular}
\end{center}
\caption{Growth rate $\omega_i$ as a function of $(\alpha,\beta)$
at neutral stability in XPPF.
(a) $Re=66.47$ and $Ro=24$,
(b) $Re=77.03$ and $Ro=1$,
(c) $Re=682.8$ and $Ro=0.05$,
(d) $Re=5776$ and $Ro=0.000587$.
The neutrally stable modes are indicated by white stars.
\label{modes_PPF}}
\end{figure}

Figure \ref{modes_PPF} shows isocontours
of the growth rate $\omega_i$ in the $(\alpha,\beta$)-plane
at neutral stability conditions in XPPF
at high to low $Ro$.
\citet{Masuda} showed similar plots for XPPF,
although only for cases with $Ro\sim O(1)$.
The isocontours are symmetric about the $\beta=0$-axis since the
modes with wavenumbers $(\alpha,\beta)$ 
and $(\alpha,-\beta)$ have the same growth rate $\omega_i$.
This symmetry can be understood by considering the effective rotation rate
$\mathbf{\Omega}^{ef}=\Omega \hat{\vetx}-(\partial U/2\partial y)\hat{\vetz}$,
where the last term is the rotation rate caused by mean shear and
$\hat{\vetz}$ is the unit vector in the $z$-direction.
In the bottom and top half of the channel, $\mathbf{\Omega}^{ef}$ 
has a negative and positive inclination angle with the $x$-axis, respectively,
which leads to the same instability on both sides of the channel,
but with opposite inclination
angles to the $x$-axis.
To illustrate this, figure \ref{vis}(a)
visualizes the vortical structure
of the critical modes in XPPF at $Re=77.02$ and $Ro=1$.
In the bottom and top half of the channel the vortical structures have
a negative and positive inclination angle to the $x$-axis
caused by modes with $\beta_c >0$ and $\beta_c < 0$, respectively.
Modes with $\beta_c >0$ and $\beta_c < 0$ also
have a larger velocity disturbance
and Reynolds shear stresses
in the bottom and top half of the channel, 
respectively, see figure \ref{vis}(c,e). 
{The velocity disturbances and Reynolds shear stresses
are obtained by averaging over $xz$-planes.
Due to streamwise rotation, all three Reynolds shear stress components
become non-zero \citep{Oberlack}.

Observations at other
$Ro$ are qualitatively similar, although
the inclination angle of the vortical structures
with the $x$-axis is smaller at lower and higher $Ro$.
When $Ro\rightarrow 0$, two-dimensional modes with $\beta=0$
become more prominent and are the most unstable modes
if $Ro$ is sufficiently small (figure \ref{modes_PPF}.d).

In contrast, the isocontours of the growth rate
$\omega_i$ in the $(\alpha,\beta$) plane
at neutral stability conditions in XPCF at four $Ro$, shown in
figure \ref{modes_PCF}, are not symmetric about the $\beta=0$ axis,
with $\omega_i$ generally being greater for $\beta>0$.
\begin{figure}
\begin{center}
\begin{tabular}{cc}
\includegraphics[width=6.5cm]{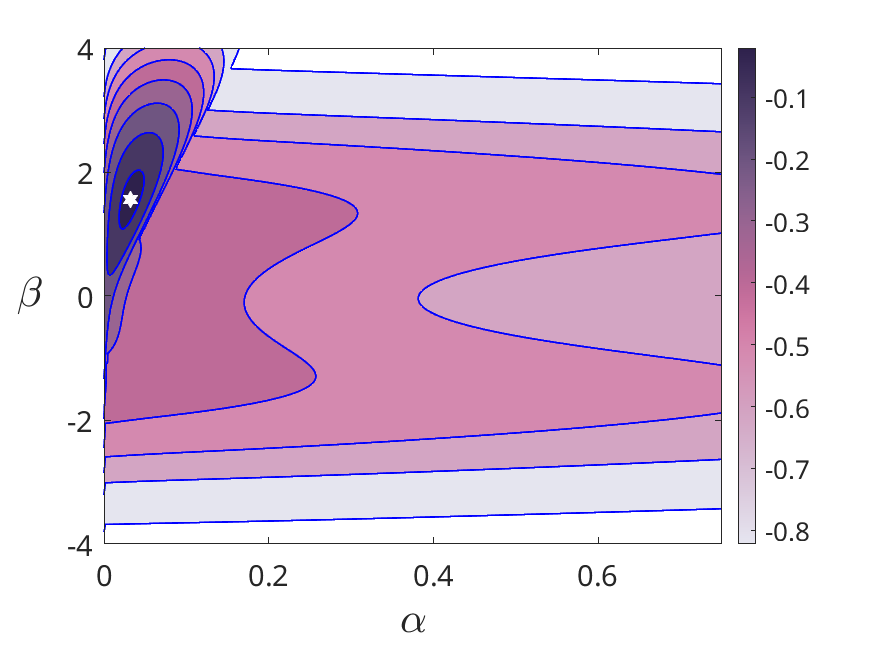}&
\includegraphics[width=6.5cm]{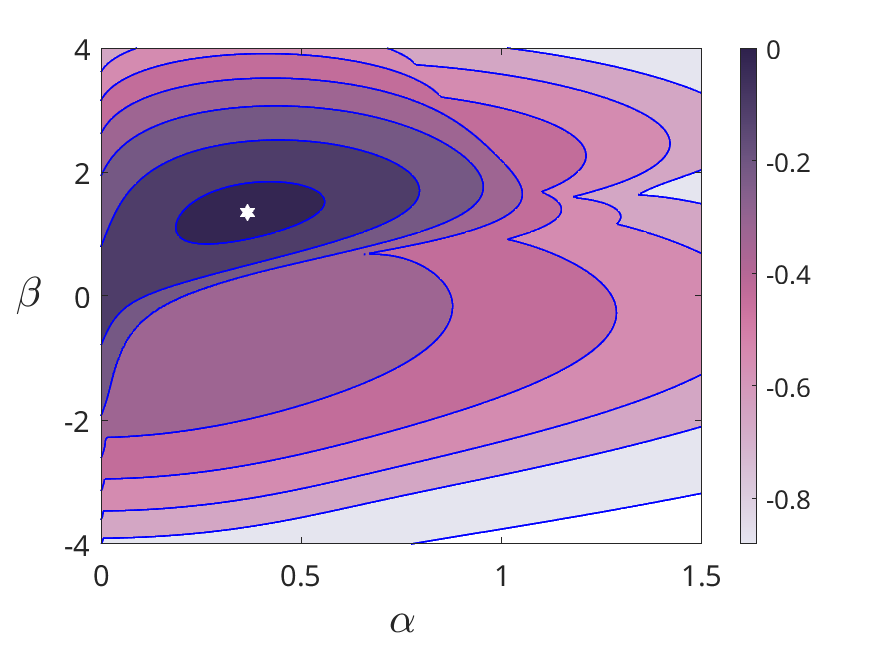}\\
(a) & (b)\\
\includegraphics[width=6.5cm]{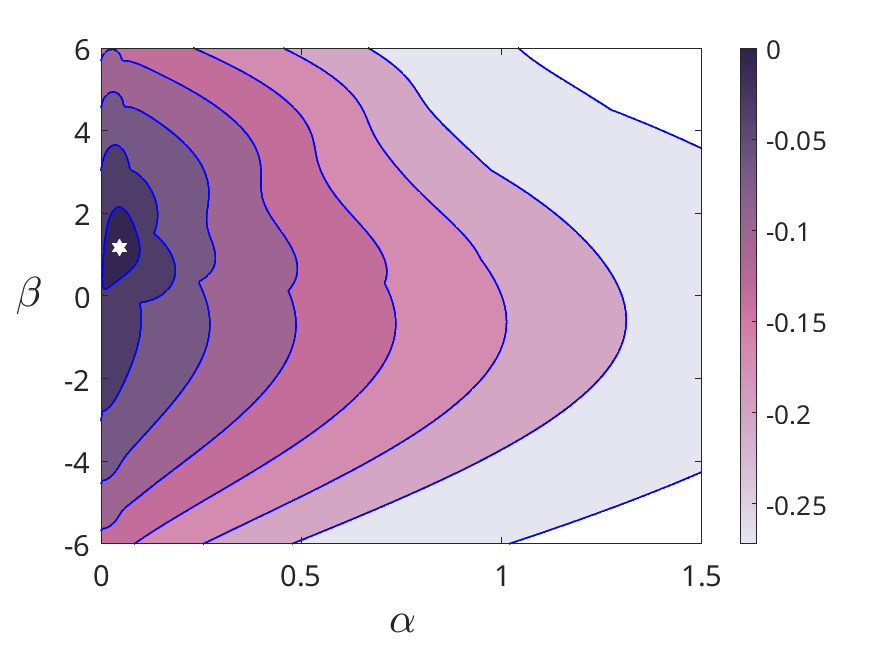}&
\includegraphics[width=6.5cm]{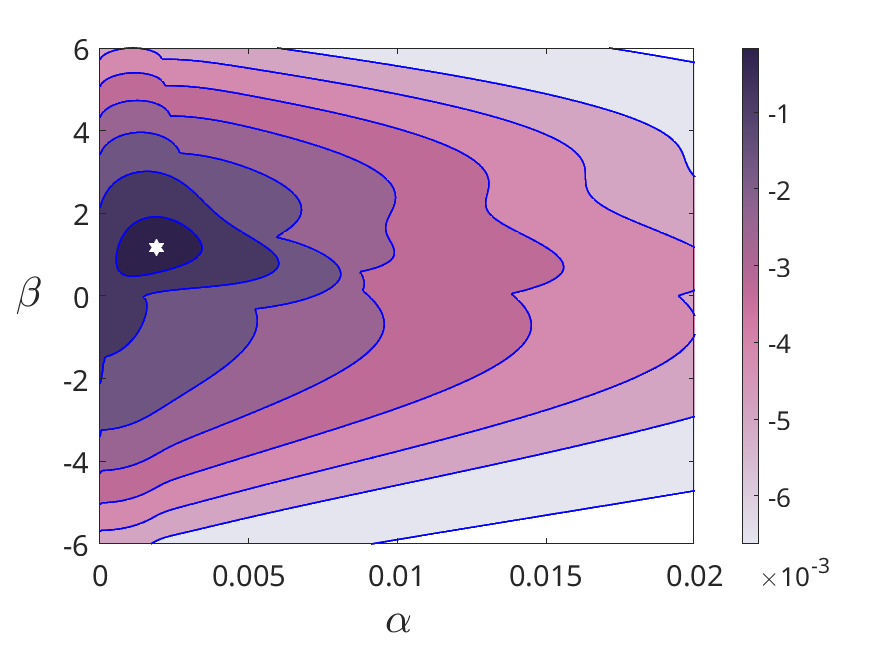}\\
(c) & (d)
\end{tabular}
\end{center}
\caption{Growth rate $\omega_i$ as a function of $(\alpha,\beta)$ 
at neutral stability in XPCF.
(a) $Re=20.68$ and $Ro=24$,
(b) $Re=28.14$ and $Ro=1$,
(c) $Re=340.9$ and $Ro=0.05$,
(d) $Re=8496$ and $Ro=0.002$.
The neutrally stable mode is indicated by a white star.
\label{modes_PCF}}
\end{figure}
Vortical structures of the critical mode with $\beta_c>0$
in XPCF at $Re=28.14$ and $Ro=1$, visualized in figure \ref{vis}.(b),
are centered in the middle of the channel and have a negative
inclination angle with the $x$-axis,
like the effective rotation rate $\mathbf{\Omega}^{ef}$. 
\begin{figure}
\begin{center}
\begin{tabular}{cc}
\includegraphics[width=6.5cm]{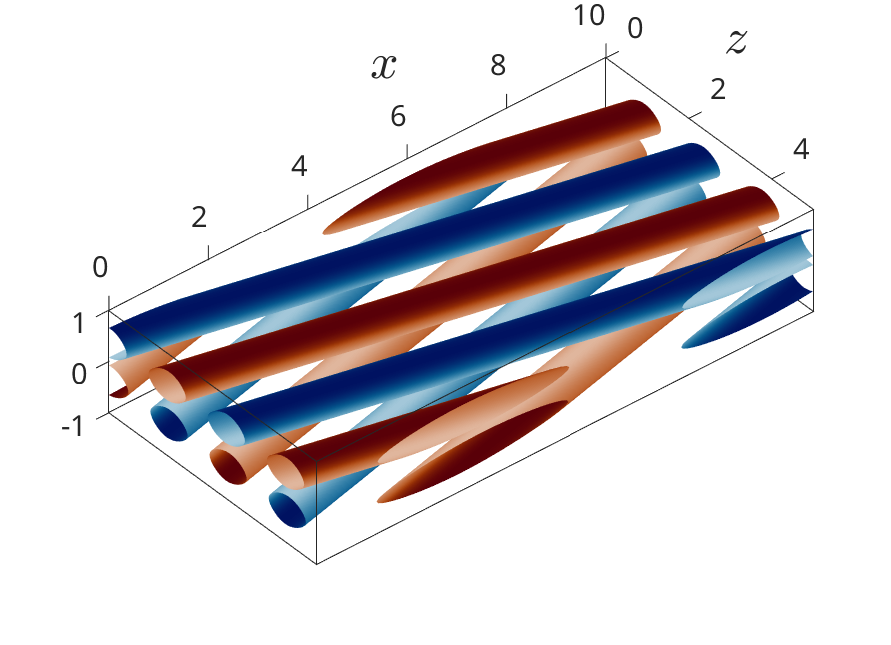}
&
\includegraphics[width=6.5cm]{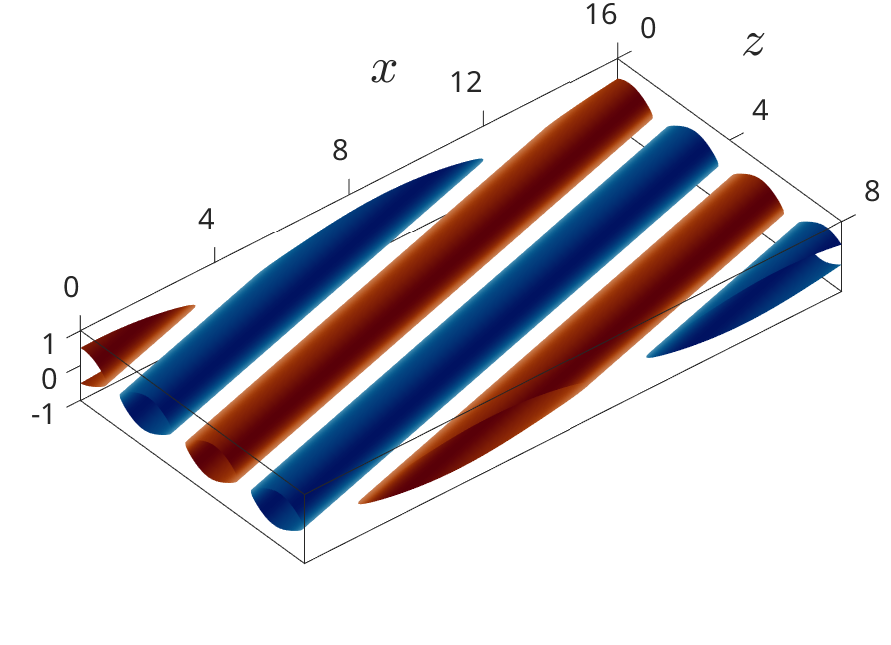}\vspace{-5mm}\\
(a) & (b) \\
\\
\\
\includegraphics[width=6.0cm]{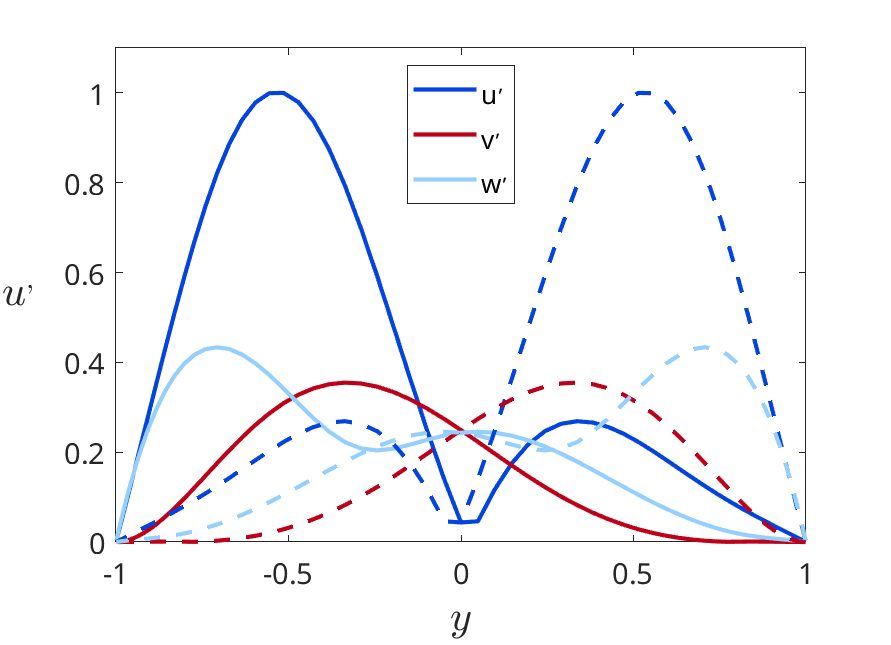}
&
\includegraphics[width=6.0cm]{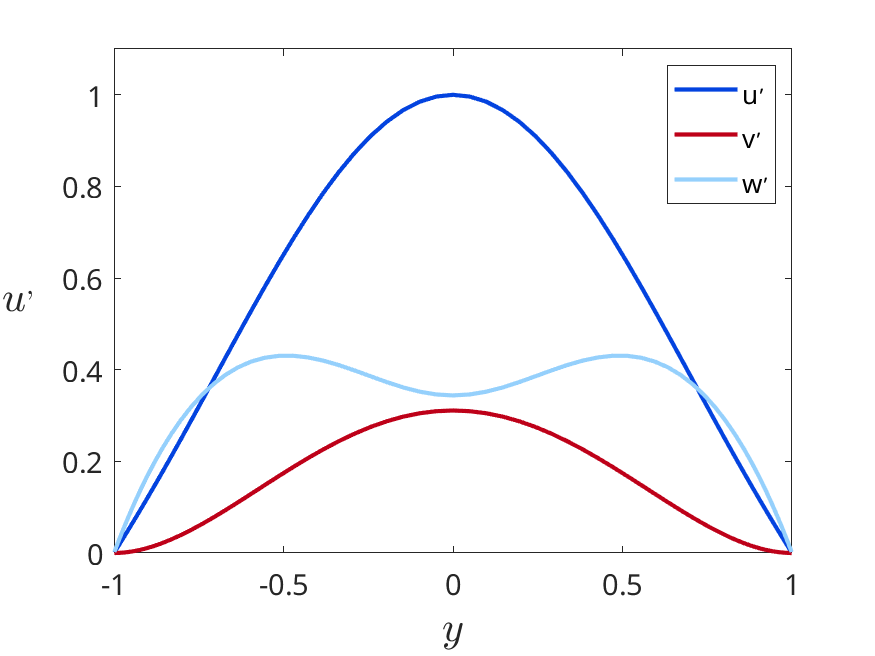}\\
(c) & (d)\\
\includegraphics[width=6.0cm]{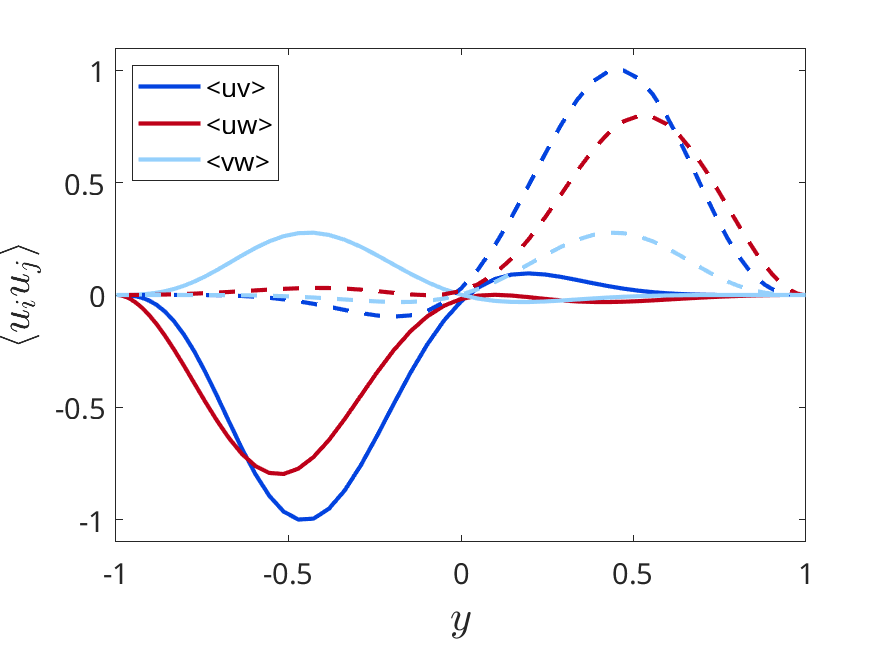}
&
\includegraphics[width=6.0cm]{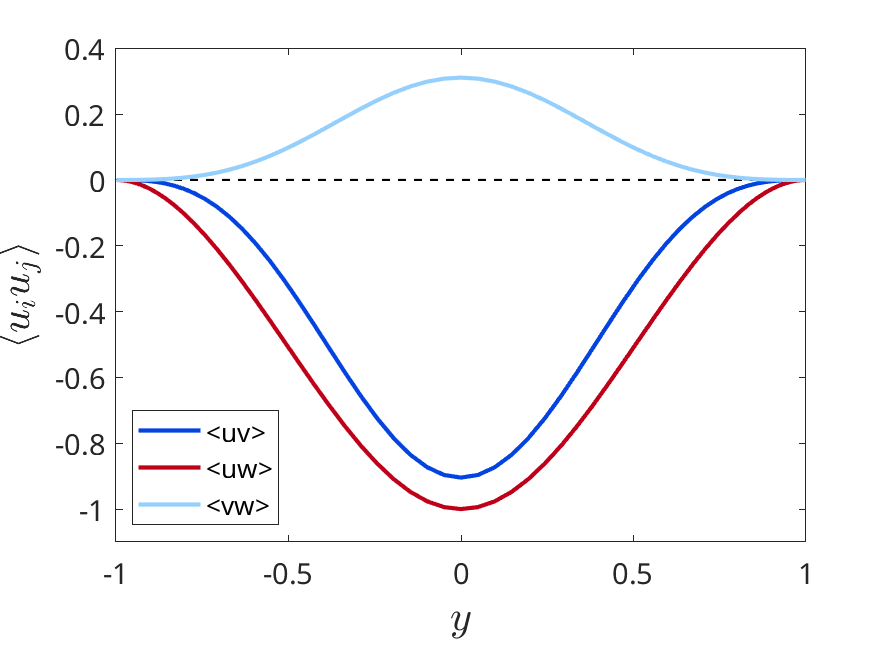}\\
(e) & (f)
\end{tabular}
\end{center}
\caption{(a) Visualization 
using the Q criterion \citep{Hunt} and
(c) root-mean-square of the velocity disturbances
and (e) the Reynolds shear stresses
of the 
two critical modes with $\beta_c >0$ and $\beta_c < 0$
shown by the solid and dashed lines, respectively,
in XPPF at $Re=77.02$ and $Ro=1$
(b) Visualization
using the Q criterion \citep{Hunt} and 
(d) root-mean-square of the velocity disturbances
and (f) the Reynolds shear stresses
of the critical mode in
XPCF at $Re=28.14$ and $Ro=1$.
The vortices in (a) and (b) are coloured by the streamwise vorticity
with blue and red denoting positive and negative values, respectively.
\label{vis}}
\end{figure}
Figure \ref{vis}(d,f) shows that this mode also has the largest
streamwise and wall-normal velocity disturbance 
and Reynolds shear stress
amplitudes in the centre of the channel. 
Observations at other $Ro$ are again quantitatively similar,
with the differences that at lower and higher $Ro$ the
inclination angle is smaller, and the wall-normal
and spanwise velocity disturbances are negligible compared
to the streamwise one if $Ro\ll 1$ (not shown here).

\section{High rotation number asymptotics}

We analyze the asymptotic behavior seen in figures \ref{neutral} and \ref{angle}.
For classical Rayleigh-B{\'e}nard convection between two horizontal
flat plates and assuming wall-normal velocity perturbations
$v(\vetx,t) = \hat{v}(y) e^{i(\alpha x + \beta z - \omega t)}$
the linearized
perturbation equation for a neutral stability mode with $\omega=0$ 
can be written as
\begin{equation}
    \left ( D^2-k^2 \right )^3 \hat{v} = - Ra\, k^2 \hat{v},
    \label{rb}
\end{equation}
where $Ra$ is the Rayleigh number \citep{Chandrasekhar}.  
The marginally stable mode is stationary with $\omega=0$.
When the problem
is nondimensionalized with the gap width $2\delta$ as length scale,
it can be shown that the critical Rayleigh number
$Ra_c=1707.762$ and wavenumber $k_c = 3.117$ \citep{Chandrasekhar}.

We now consider ZPPF and ZPCF. In these cases, the eigenvalue
problem of the LSA is the same as of XPPF and XPCF given
by equations (\ref{eigen}), except that the rotation term
$\alpha Ro$ is replaced by $\beta Ro$ since
the system rotation is about the $z$-direction, that is,
$\mathcal{L}_R=i\beta Ro$ and
$\mathcal{L}_C = i\beta(U'-Ro)$.
Since the most unstable mode
is two-dimensional with $\alpha=0$ in ZPPF and ZPCF \citep{Lezius,Wall},
and the neutral stability mode is stationary with $\omega=0$,
we can derive, from the eigenvalue problem (\ref{eigen})
after substituting $\mathcal{L}_R$ and $\mathcal{L}_C$ that 
\begin{equation}
    \left ( D^2-k^2 \right )^3 \hat{v} = - Re^2_d Ro(U'-Ro)\, \beta^2 \hat{v},
    \label{rotz}
\end{equation}
by eliminating $\hat{\eta}$ from the eigenvalue problem.
The boundary conditions for $\hat{v}$ are the same as in
Rayleigh-B{\'e}nard convection.
The perturbation equation for PCF is nondimensionalized
using the gap width $2\delta$ and the velocity difference between the walls
$2U_w$, to retain the similarity with the 
Rayleigh-B{\'e}nard convection stability problem, so that
$Re_d = 4U_w \delta/\nu = 4 Re$. For PPF, we keep $Re_d=Re$.
Further, $U'=1$ for PCF.
The similarity between perturbation equations
(\ref{rb}) for Rayleigh-B{\'e}nard convection
and (\ref{rotz}) for ZPCF then leads to
$16Re^2_c Ro(1-Ro)=Ra_c$ and $\beta_c = 3.117/2 = 1.558$ if
$\delta$ is used as length scale \citep{Lezius}.
Consequently, the minimum $Re_c = \sqrt{Ra_c}/2=20.6625$ 
in ZPCF occurs at $Ro=1/2$. 
In this case, $Re_c$ and $Re_E$ coincide,
as shown by \cite{Joseph1970,Busse1970}, similar to Rayleigh–B{\'e}nard flow.

We return to XPPF and XPCF and apply the same procedure. 
In XPPF the neutrally stable modes are not stationary,
therefore, only $\omega_i = 0$ and $\omega = \omega_r$
with $\omega_r$ the (real) wave frequency.
Considering neutrally stable modes with $\omega = \omega_r$ and
eliminating $\hat{\eta}$ from the eigenvalue problem (\ref{eigen}),
we find that 
\begin{equation}
\begin{split}
    \left ( D^2-k^2 \right )^3 \hat{v} = 
    &- Re^2_d \alpha Ro \left (\beta U'-\alpha Ro \right ) \hat{v} \\
    &+ Re^2_d \alpha^2 \mathcal{U}_X \mathcal{L}_X \hat{v} \\
    &+ i Re_d \alpha \left [ \mathcal{U}_X \left (D^2-k^2 \right )^2
    + \left (D^2 - k^2 \right ) \mathcal{L}_X \right ] \hat{v},
    \end{split}
    \label{rotx}
\end{equation}
where $\mathcal{U}_X = U - \omega_r/\alpha$ and
$\mathcal{L}_X = \mathcal{U}_X \left (D^2-k^2 \right) - U''$.  
Note that the most unstable modes are three-dimensional.
In XPCF, $U''=0$ and $\mathcal{U}_X = U$ since
the neutrally stable modes are stationary ($\omega_r=0$) if
the walls move with the same speed but in opposite directions.

Of all three terms on the right-hand-side
of equation (\ref{rotx}) only the first
contains $Ro$ and $Ro^2$ and therefore dominates if $Ro \gg 1$.
This has been verified
by comparing the terms using the eigenvalue solver for the LSA.
In that case, when $Ro \gg 1$ and only the first
term is relevant, equations
(\ref{rotz}) and (\ref{rotx})
are equivalent if $\alpha Ro$ in equation
(\ref{rotx}) for XPPF/XPCF is equal to
$\beta Ro$ in equation (\ref{rotz}) for ZPPF/ZPCF.
That is, the perturbation equations are similar when the component of $\vetk$ parallel to the rotation axis, multiplied by the rotation rate, is the same in the streamwise and spanwise rotating cases. This implies that the Coriolis force acting on a slightly oblique mode in a rapidly streamwise rotating flow can have the same effect on the wall-normal velocity perturbation as the Coriolis force acting on a purely streamwise (longitudinal) mode in a spanwise rotating flow.
We know that the minimum critical $Re_c$ in
ZPPF and ZPCF occurs at $Ro^c_{ZPPF}=0.3366$ and $Ro^c_{ZPCF}=0.5$,
respectively \citep{Lezius,Wall}.
The similarity of the perturbation equations
when $\alpha Ro$ in XPPF/XPCF is equal to $\beta Ro$ in ZPPF/ZPCF
means that $\beta_c$ and $Re_c$ in the streamwise rotating cases
are the same as $\beta_c$ and minimum $Re_c$ in the spanwise rotating cases.
Furthermore, $Re_c$ in XPPF and XPCF
is found for that $\theta$ when
$Ro \tan \theta = Ro^c_{ZPPF}=0.3366$ and $Ro \tan \theta = Ro^c_{ZPCF}=0.5$,
respectively, where $\theta = \arctan(\alpha / \beta)$ is again
the angle of $\vetk$ with the $z$-axis in the streamwise rotating case. 
We can approximate
$\tan \theta \simeq \theta$ when $Ro\gg 1$, so that
the critical mode in
XPPF and XPCF obeys $\theta = 0.3366/Ro$ and
$\theta = 0.5/Ro$, respectively.
With $Ro \tan \theta = Ro^c_{ZPCF}=0.5$ 
equation (\ref{rotx}) for XPCF becomes
\begin{equation}
    \left ( D^2-k^2 \right )^3 \hat{v} = - \frac{1}{4} Re^2_d \beta^2 \hat{v}.
    \label{pert_XPCF}
\end{equation}
The similarity between equation (\ref{pert_XPCF})
and equation (\ref{rb}) for Rayleigh-B{\'e}nard convection
gives $\beta_c = 1.558$ and $Re^2_d/4 = 4 Re^2 = Ra$, therefore,
$Re_c = \sqrt{Ra_c}/2 = 20.6625$ in XPCF when $Ro \rightarrow \infty$.

The results of these considerations,
$Re_c=20.6625$, $\beta_c=1.558$ and $\theta_c=0.5/Ro$ in XPCF, and
$Re_c=66.45$, $\beta_c=2.459$ and $\theta_c=0.3366/Ro$ in XPPF,
are shown by dashed lines in figures \ref{neutral}
and \ref{angle}(b,c), confirming that these
values are approached for $Ro\gg 1$.

In summary, the critical Reynolds number $Re_c$ and
wavenumber $\beta_c$ in XPPF and XPCF
become independent of $Ro$ and approach the minimum $Re_c$
and corresponding $\beta_c$ in ZPPF and ZPCF, respectively, 
for $Ro\rightarrow \infty$. Moreover, 
the linear stability of ZPCF as well as
XPCF at $Ro\rightarrow \infty$ share similarities
with that of Rayleigh-B{\'e}nard convection.
For $Ro \lesssim 5$ in the streamwise rotating cases,
the remaining terms on the right-hand-side of equation (\ref{rotx})
become significant, and the similarity with the spanwise rotating cases
is lost.

In XPCF, the critical Reynolds number for energy instability $Re_E$ 
is identical to that in NPCF because energy stability is unaffected by rotation \citep{Joseph1970,Joseph1976}; the Coriolis term vanishes in the energy equation. 
In this case, the eigenvalue problem for energy instability is also equivalent
to that of the LSA for Rayleigh-B{\'e}nard convection 
given by equation (\ref{rb}) \citep{Joseph1966},
yielding $Re_E=\sqrt{Ra_c}/2=20.66$ for both NPCF and XPCF \citep{Busse1970,Joseph1970,Reddy,Barletta}.
The present analysis shows that $Re_c$ converges to this same value
in the limit $Ro\rightarrow\infty$, showing
that linear and energy stability coincide,
ruling out subcritical transition.
\cite{Busse1970} demonstrated
the same result for ZPCF at $Ro=0.5$
and noted its extension to XPCF as $Ro\rightarrow\infty$.
\cite{Joseph1970,Joseph1976}, using a different approach within the framework of spiral flow between concentric cylinders, confirmed the coincidence of energy and linear stability in XPCF at $Ro\rightarrow\infty$ 
for $\alpha\rightarrow 0$, $\theta Ro=0.5$ and $\beta=1.558$, consistent
with the present results.
This mode is the most susceptible to transient growth in NPCF due to the non-normality of the linearized Navier–Stokes operator \citep{Reddy}.

Thus, streamwise rotation, like spanwise anti-cyclonic rotation, preferentially destabilizes the mode showing maximal transient growth without rotation, explaining the strong destabilizing effect of rotation and making the linearized Navier–Stokes operator effectively normal again. A similar argument applies to the XPPF case; further details can be found in the study by \citet{Jose2020}.

In the asymptotic limit $Ro\rightarrow 0$ of XPPF
and XPCF the critical vortices also align with
the $x$-axis and thus $\vetk_c$ aligns with the $z$-axis,
giving $\alpha = \beta \tan \theta \simeq \beta \theta$.
The first term on the right hand side of equation
(\ref{rotx}) then approaches $- Re^2_d \theta Ro U'\, \beta^2 \hat{v}$
since $\alpha Ro \ll \beta U'$.
When $\theta \propto Ro$, $Re_d \propto 1/Ro$ and $\beta$ is constant,
all three terms on the right-hand-side of
perturbation equation (\ref{rotx}) remain constant and significant.
This behaviour, $\theta_c \propto Ro$ and $Re_c \propto 1/Ro$
(noting that $Re_c \propto Re_d$)
is observed in figures \ref{neutral}
and \ref{angle}(c) in XPPF and XPCF
in the limit $Ro\rightarrow 0$.

\section{Results: direct numerical simulations}

Subcritical transition to turbulence can occur in NPPF and NPCF,
resulting in stable coexisting
laminar and turbulent states \citep{Grossmann,Manneville}.
However, when $Re$ is gradually reduced,
uniformly turbulent NPPF and NPCF
become transitional before relaminarizing,
and turbulent-laminar flow patterns develop if
the flow domain is sufficiently large \citep{Shimuzu,Tuckerman}. 
Subcritical transition and
transitional regimes have also been observed in ZPCF,
at higher $Re$ than in NPCF, when the rotation is cyclonic
and $Re_c \rightarrow \infty$ \citep{Tsukahara,Brethouwer2012},
but not yet when the rotation is anticyclonic and
destabilizes the flow \citep{Alfredsson,Tsukahara}.
Instead, regular and steady streamwise vortices are observed
at $Re$ near $Re_c$.
In ZPPF subcritical transition has only been observed
at very low $Ro$ \citep{Jose2017}, and
turbulent-laminar patterns only appear in some $Re-Ro$ range
on the channel side stabilized by rotation
\citep{Brethouwer2017,Brethouwer2019}.

It is not yet known whether a subcritical
transition can occur, and whether the transitional regime exists
in XPPF and XPCF. 
We carry out DNS of XPPF and XPCF to address these questions.
The DNS cover the range
$Re \leq 2000$ in XPPF and $Re \leq 1000$ in XPCF,
and $0 \leq Ro \leq 0.8$, and also include
non-rotating and rapidly rotating cases at $Ro=24$.
A computational domain $L_x/\delta \times L_z/\delta$ of $110 \times 50$ and
$250 \times 125$ is used in the DNS of XPPF and XPCF, respectively, 
and a spatial resolution of $512 \times 33 \times 512$ in the
streamwise, wall-normal and spanwise direction, unless
otherwise specified. 
These computational domain sizes are similar to those used
in DNS by \citet{Brethouwer2012} and large enough
to accommodate large-scale turbulent-laminar patterns.
The resolution in the XPCF cases is finer than that
determined by \citet{Manneville_Rolland} for relatively
well-resolved uniformly turbulent and transitional regimes in NPCF.
The formation of turbulent-laminar patterns in NPCF
is not very sensitive to resolution, with only a gradual
downward shift of the $Re$ threshold for patterns
when resolution becomes coarse \citep{Manneville_Rolland}.

To determine the lower $Re$ threshold of sustained vortices or turbulence
at a given $Ro$, we initialize the DNS
with turbulent flow at sufficiently high $Re$ and
reduce $Re$ in small steps until the flow becomes laminar.
The lower threshold for sustained turbulence,
called $Re_t$, is defined as the lowest $Re$
at which turbulence or turbulent patterns persist
for a time period of at least $2\cdot10^5(\delta / U_{cl,w})$ in our DNS.
This does not preclude that
turbulence or turbulent patterns
eventually disappear on longer time scales.


Firstly, we discuss DNS results of PPF
using visualizations of the instantaneous streamwise
velocity field in an $xz$-plane near the wall at $y=-0.9$
shown in figure \ref{patterns_ppf}.
\begin{figure}
\begin{center}
\begin{tabular}{cc}
\includegraphics[width=5.5cm]{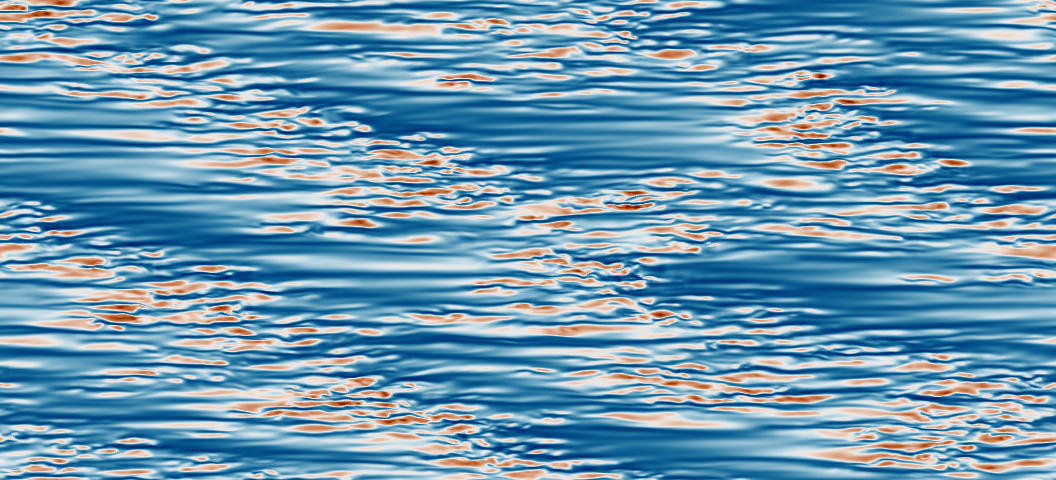}&
\includegraphics[width=5.5cm]{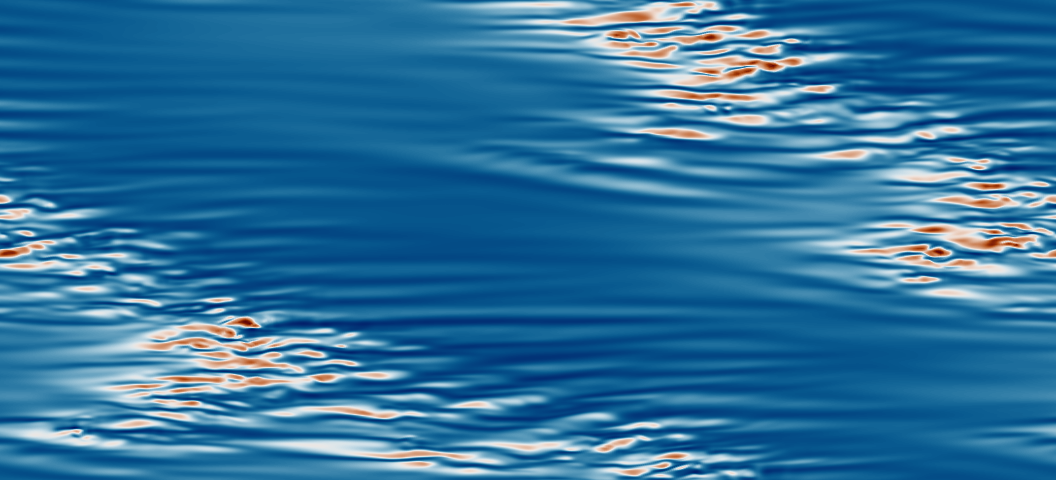} \\
(a) $Ro=0$, $Re=1400$ & (b) $Ro=0$, $Re=1000$ \\
\includegraphics[width=5.5cm]{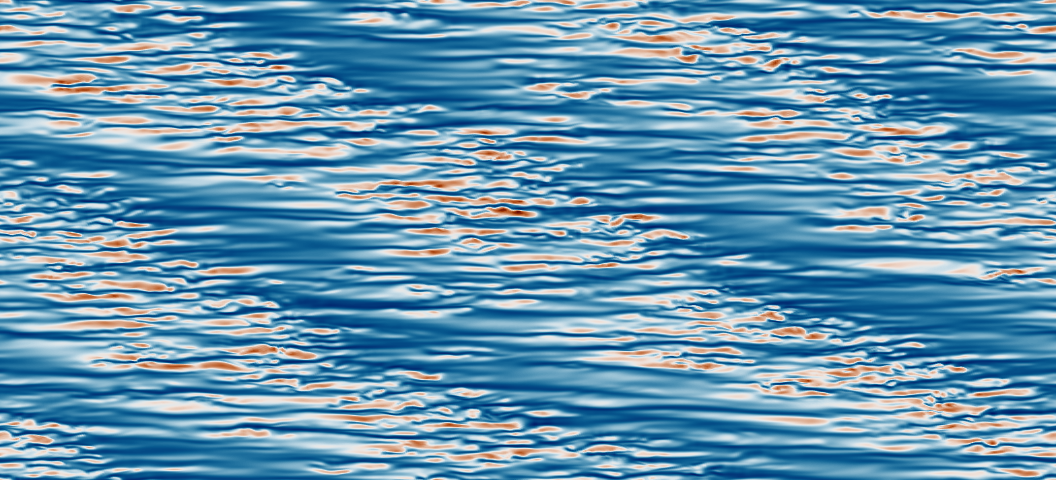}&
\includegraphics[width=5.5cm]{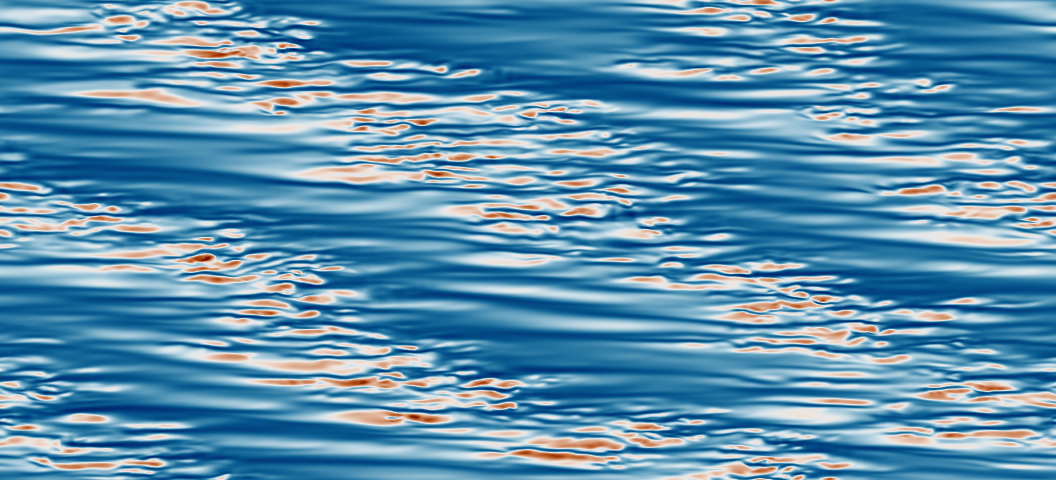} \\
(c) $Ro=0.025$, $Re=1400$ & (d) $Ro=0.04$, $Re=1200$ \\
\includegraphics[width=5.5cm]{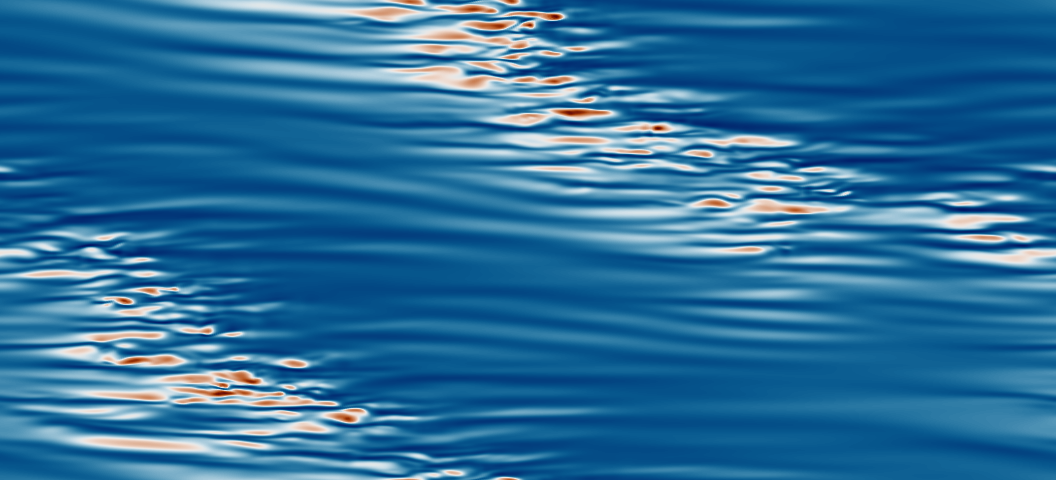}&
\includegraphics[width=5.5cm]{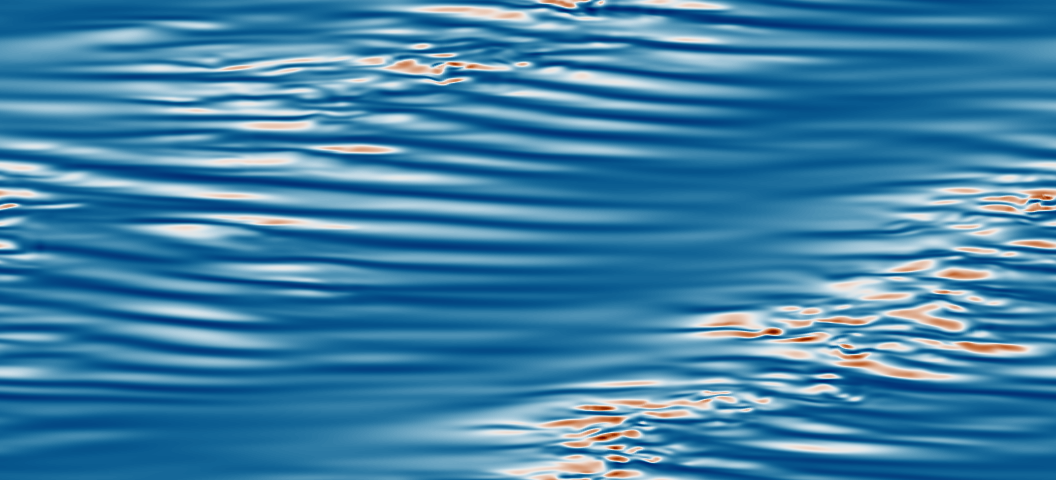} \\
(e) $Ro=0.0356$, $Re=1000$ & (f) $Ro=0.05$, $Re=950$\\
\includegraphics[width=5.5cm]{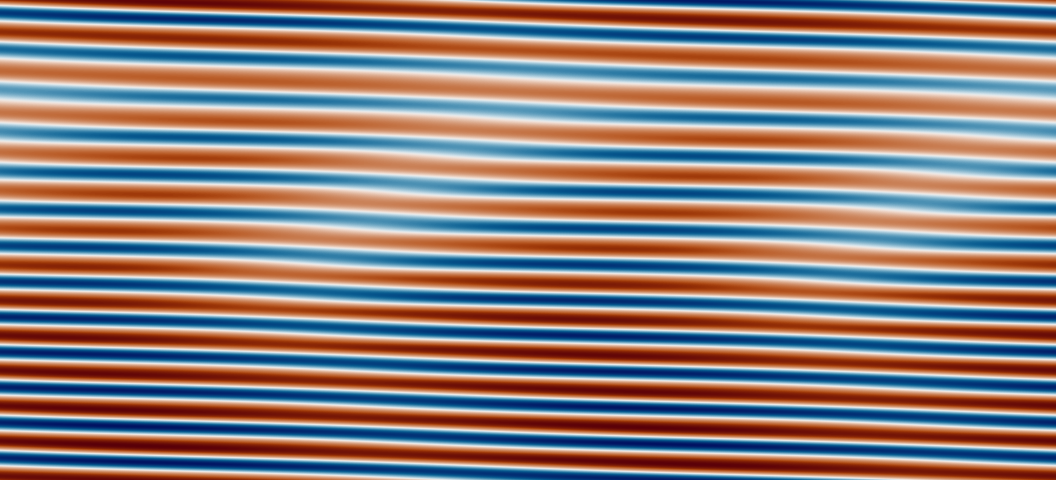}&
\includegraphics[width=5.5cm]{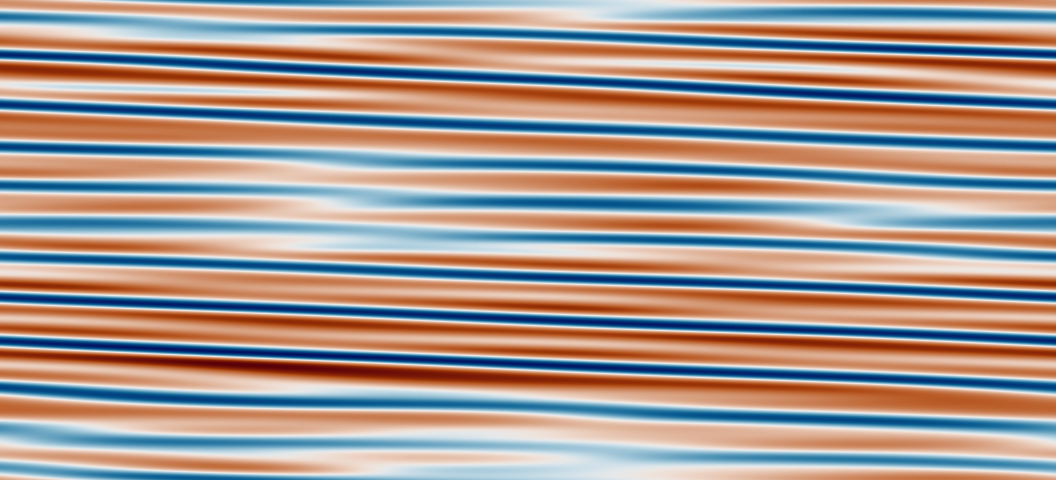} \\
(g) $Ro=0.04$, $Re=950$ & (h) $Ro=0.05$, $Re=900$\\
\includegraphics[width=5.5cm]{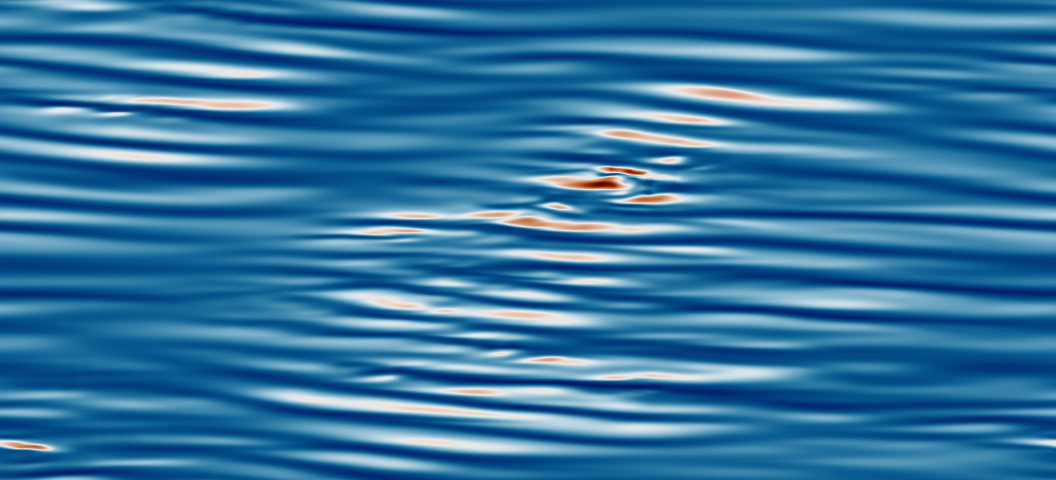}&
\includegraphics[width=5.5cm]{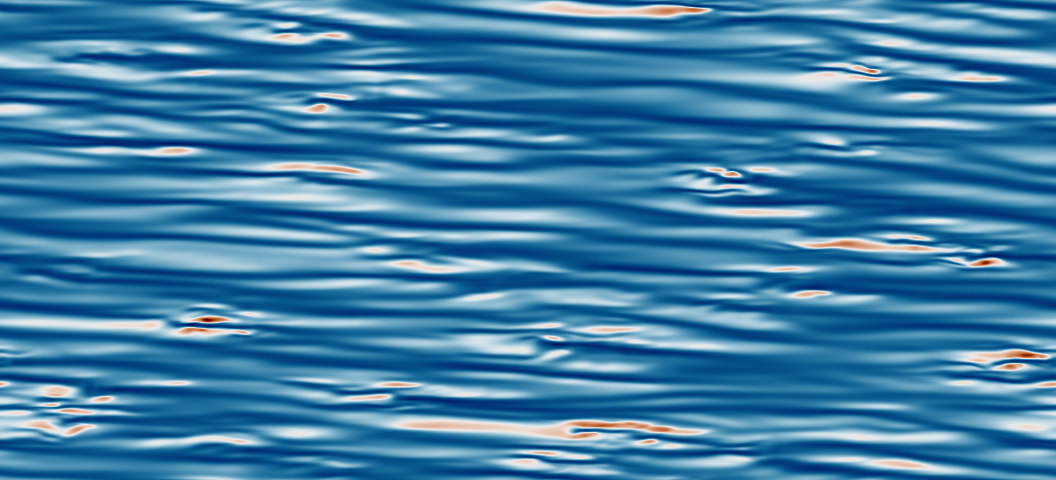} \\
(i) $Ro=0.07$, $Re=900$ & (j) $Ro=0.1$, $Re=900$\\
\includegraphics[width=5.5cm]{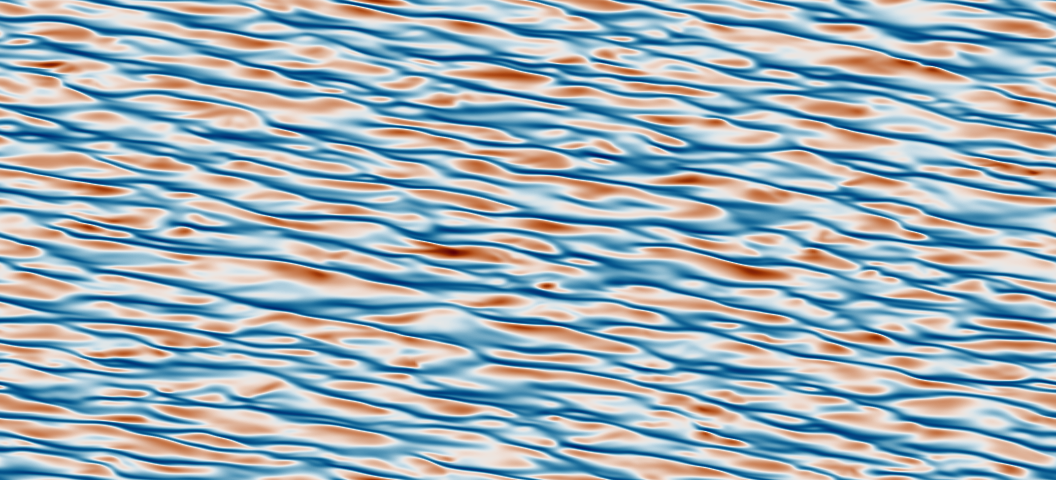}&
\includegraphics[width=5.5cm]{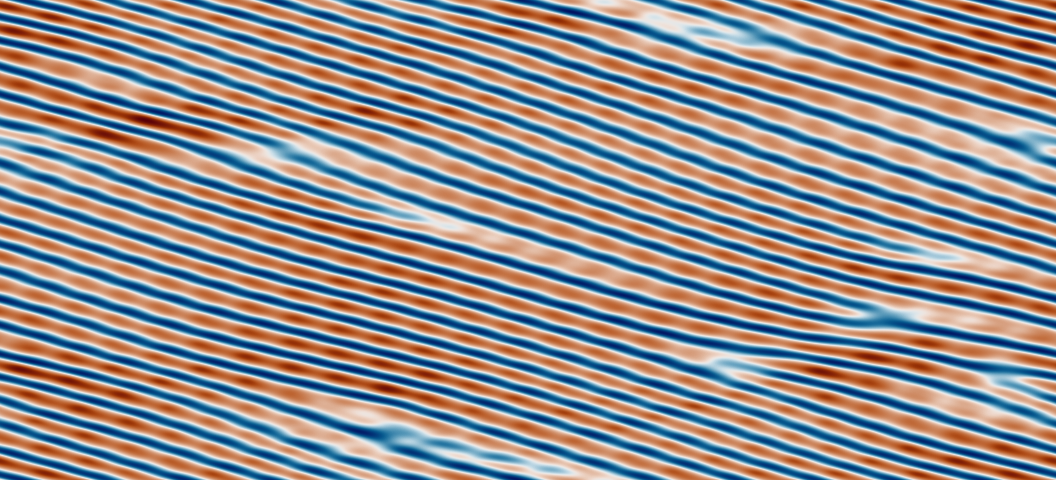} \\
(k) $Ro=0.4$, $Re=700$ & (l) $Ro=0.8$, $Re=110$
\end{tabular}
\end{center}
\caption{Visualization of the streamwise velocity field in XPPF in an
$xz$-plane at $y=-0.9$.
\label{patterns_ppf}}
\end{figure}
Additional visualizations of the velocity field are
presented in the Supplementary Material.
Hereafter, $Re= U_{cl}\delta/\nu$ for PPF, where $U_{cl}$ is the centreline
velocity of the laminar base flow. The bulk Reynolds number
$Re_b=U_b \delta/\nu = 2 Re/3$, where $U_b$ is the bulk mean velocity.
We observe in NPPF ($Ro=0$) uniform turbulence
at $Re=2000$, and transitional turbulence with
oblique turbulent-laminar patterns at $Re=1400$ (figure \ref{patterns_ppf}.a). 
When $Re$ decreases, the turbulent patterns become less
structured and at $Re_t=1000$ one oblique turbulent
band persists in a laminar-like flow
environment (figure \ref{patterns_ppf}.b),
while below $Re_t$ the flow relaminarizes.

These results for NPPF are broadly consistent with those
of \citet{Shimuzu}.
They observed local relaminarization at $Re \approx 1800$
and turbulent patterns at lower $Re$ till about $800$ in NPPF.
This $Re$ threshold for turbulent patterns is lower
than in our DNS, which may be a result of the larger
computational domain in their study,
different simulation time period and
other flow forcing (constant pressure
gradient in their study vs. constant mass flow
in our study). However, using a larger computational
domain in our DNS is prohibitively expensive when
covering a wide range $Ro$, which requires many simulations.

In XPPF we also observe at low $Ro \lesssim 0.05$
a transitional regime with sustained turbulent-laminar patterns
(figure \ref{patterns_ppf}.c,d),
sometimes forming oblique bands,
at low $Re$ until $Re_t=1000$ at $Ro \leq 0.04$ and $Re_t=950$
at $Ro=0.05$ (figure \ref{patterns_ppf}.e,f).
The observed patterns span the whole channel gap width, as in NPPF,
but in the present configuration we observe differences in the DNS at low $Ro$.
At $Ro=0$, $0.025$ and $0.0356$ the flow
relaminarizes if $Re<Re_t$, while at $Ro=0.04$
and $0.05$ the turbulent patterns disappear if
$Re < Re_t$, but the flow does not relaminarize
since $Re_t > Re_c$. Instead, we observe
regular vortices nearly
aligned with the streamwise direction
without signs of turbulence
(figure \ref{patterns_ppf}.g,h).
When $Re$ is further reduced
the flow only relaminarizes once $Re \leq Re_c$.
At $Ro=0.07$ we observe spotty turbulent structures
at low $Re$ until $Re \approx 900$ 
(figure \ref{patterns_ppf}.i), and more regular
vortices at lower $Re$ until $Re_c$ when the flow relaminarizes.
When $Ro$ increases, the spotty structures gradually disappear
and turbulence becomes more uniform
(Figure \ref{patterns_ppf}.j,k).
The flow becomes less turbulent when $Re$
approaches $Re_c$ (figure \ref{patterns_ppf}.l) 
and fully relaminarizes when $Re<Re_c$.

We now study XPCF using
visualizations of the instantaneous streamwise
velocity field in an $xz$-plane at the centre at $y=0$
shown in figure \ref{patterns_pcf}.
\begin{figure}
\begin{center}
\begin{tabular}{cc}
\includegraphics[width=5.0cm]{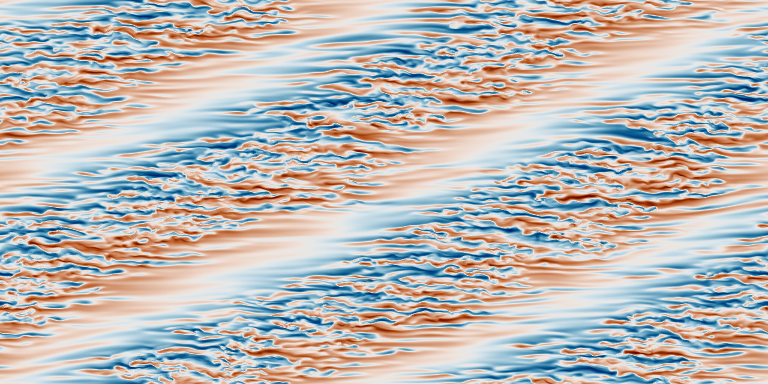}&
\includegraphics[width=5.0cm]{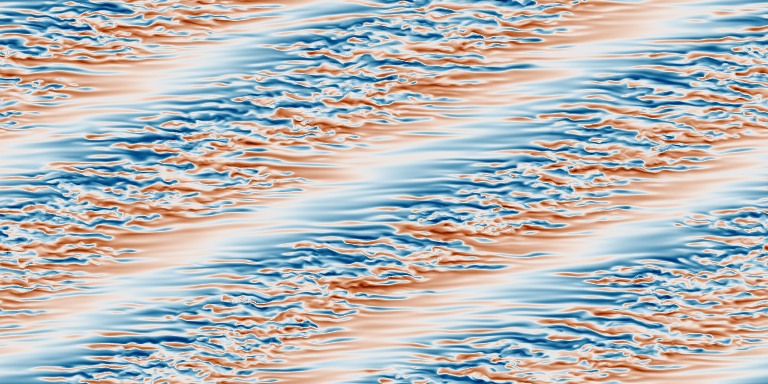} \\
(a) $Ro=0$, $Re=340$ & (b) $Ro=0.025$, $Re340$\\
\includegraphics[width=5.0cm]{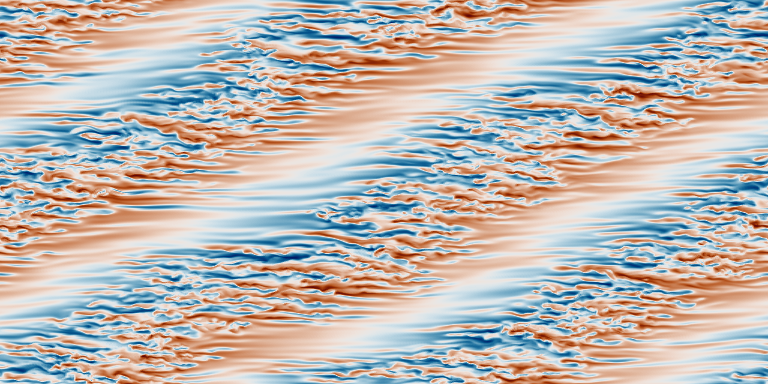}&
\includegraphics[width=5.0cm]{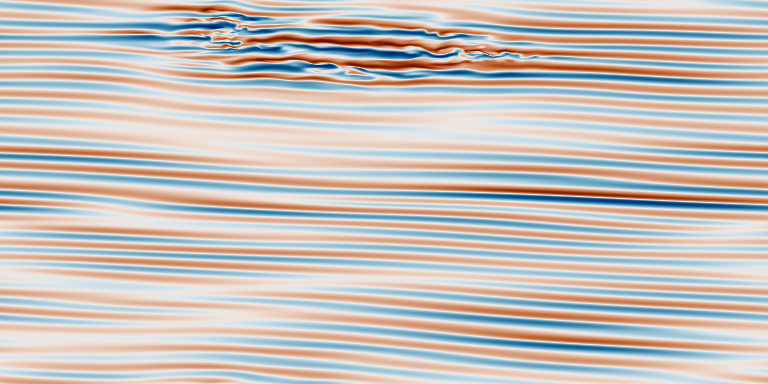} \\
(c) $Ro=0.07$, $Re=330$ & (d) $Ro=0.07$, $Re=300$\\
\includegraphics[width=5.0cm]{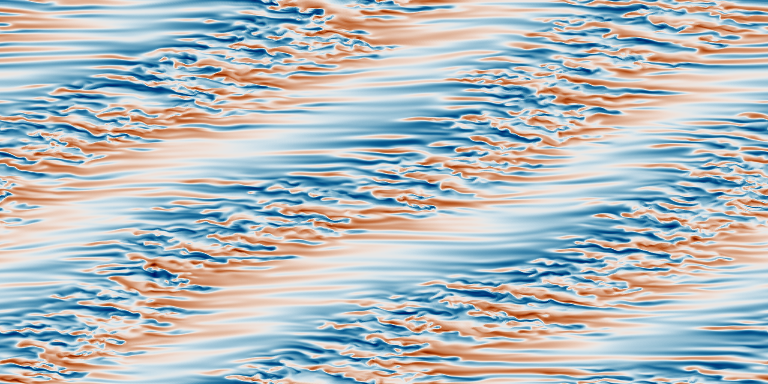}&
\includegraphics[width=5.0cm]{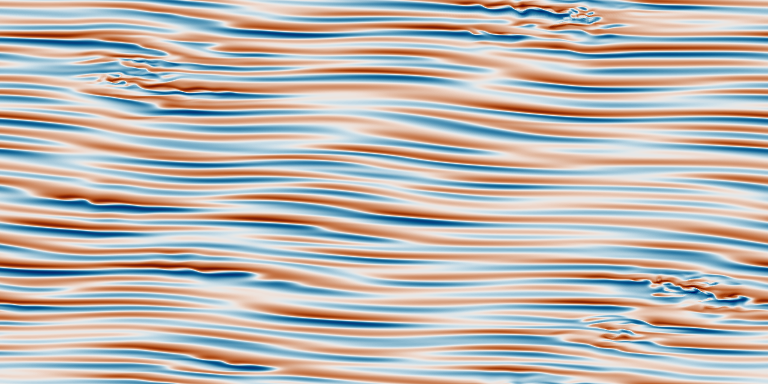} \\
(e) $Ro=0.1$, $Re=310$ & (f) $Ro=0.1$, $Re=290$\\
\includegraphics[width=5.0cm]{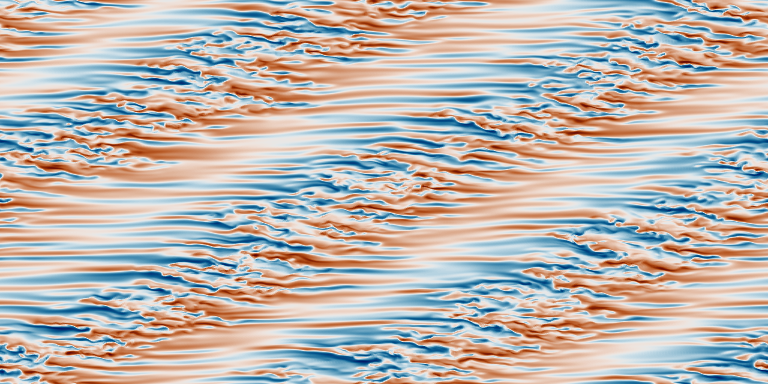}&
\includegraphics[width=5.0cm]{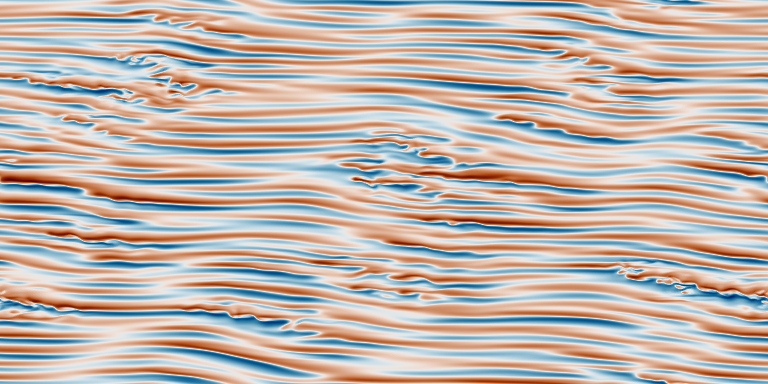} \\
(g) $Ro=0.14$, $Re=300$ & (h) $Ro=0.14$, $Re=250$\\
\includegraphics[width=5.0cm]{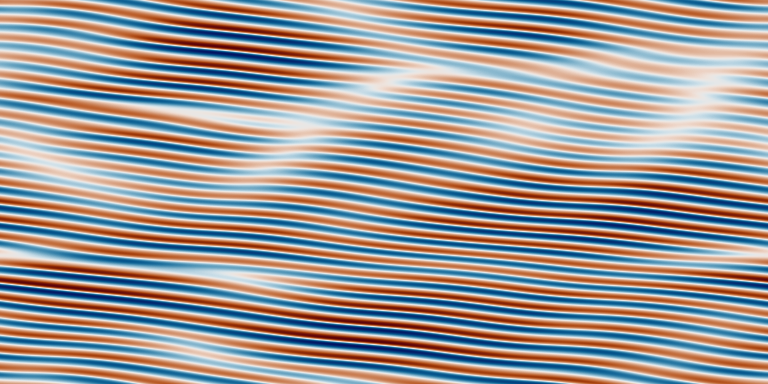}&
\includegraphics[width=5.0cm]{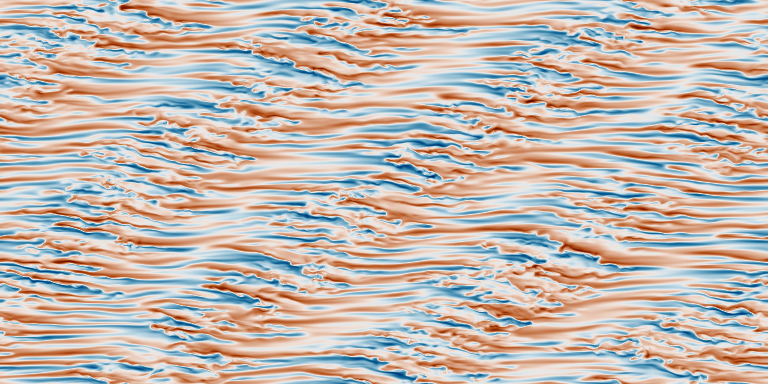} \\
(i) $Ro=0.14$, $Re=135$ & (j) $Ro=0.2$, $Re=250$ \\
\includegraphics[width=5.0cm]{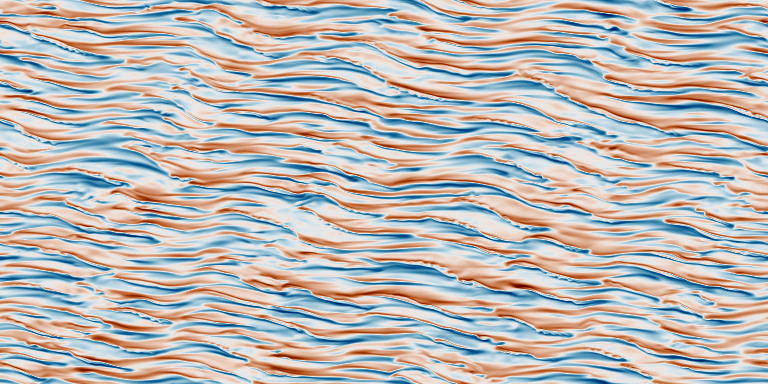}&
\includegraphics[width=5.0cm]{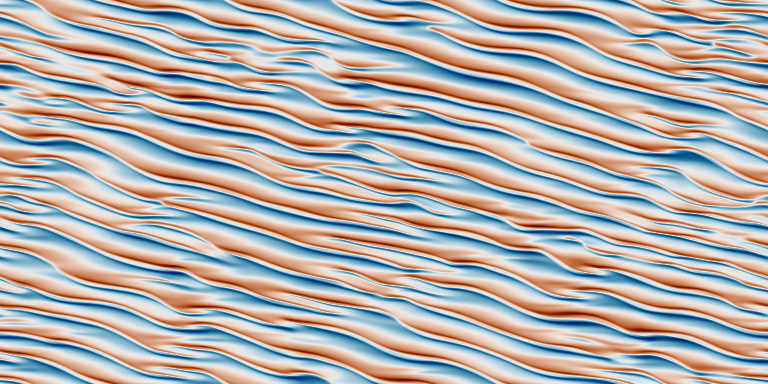} \\
(k) $Ro=0.4$, $Re=250$ & (l) $Ro=0.8$, $Re=100$
\end{tabular}
\end{center}
\caption{Visualization of the streamwise velocity field in XPCF in an
$xz$-plane at $y=0$.
\label{patterns_pcf}}
\end{figure}
Additional visualizations of the velocity field are
again presented in the Supplementary Material.
The behaviour of XPCF is qualitatively similar to that of XPPF.
In DNS of NPCF ($Ro=0$) we observe uniform turbulence
at $Re > 400$, local relaminarization
at $Re \simeq 400$, and
turbulent-laminar patterns at lower $Re$,
which are sustained
until $Re_t=340$ (figure \ref{patterns_pcf}.a).
The patterns form clearer structured oblique bands
than in NPPF. These observations are consistent
with previous studies \citep{Prigent,Duguet}, although
in DNS by \citet{Duguet} turbulent-laminar patterns
could also be sustained at somewhat lower $Re \simeq 324$.
This may be caused by a difference
in the computational domain size and simulation time
period, which was $2\cdot 10^4 (\delta/U_w)$
in the DNS by \citet{Duguet}. In our DNS, 
turbulent patterns persist for such a time period at $Re=330$,
but after a time period of nearly $10^5(\delta/U_w)$ the flow
relaminarizes.

Observations in DNS of XPCF at $Ro=0.025$, $0.05$,
$0.07$ and $0.1$ are similar.
We observe uniform turbulence at $Re > 400$,
local flow relaminarization at $Re \simeq 400$, 
and turbulent patterns and
oblique bands develop when $Re$ is gradually reduced
(figure \ref{patterns_pcf}.b, c, e).
Full relaminarization of the flow happens when $Re< 340$ at $Ro=0.025$ and
$Re< 330$ at $Ro=0.05$.
The oblique bands span the whole channel gap width, as in NPCF
and ZPCF at low cyclonic rotation rates \citep{Brethouwer2012}.
At $Ro=0.07$ and $0.1$ the turbulent pattern disappears when $Re < 310$
and $Re < 300$, respectively,
but the flow does not relaminarize when $Re$ is reduced
as long as $Re > Re_c$,
since regular vortices persist with
localized disturbances but without
larger turbulent patterns (Figure \ref{patterns_pcf}.d,f).
The flow relaminarizes once $Re < Re_c$.

Oblique band-like structures
appear in XPCF at $Ro=0.14$ if $Re\lesssim 450$.
These bands become more distinct when
$Re$ is further lowered
(figure \ref{patterns_pcf}.g), but between the turbulent bands
we see streamwise vortices and not the
clear laminar-like flow regions, as at lower $Ro$.
The turbulent bands disappear when $Re < 280$.
Localized disturbances and vortical motions persist
at $Re$ near $Re_t$ (figure \ref{patterns_pcf}.h),
while only streamwise vortices
persist at lower $Re$ (figure \ref{patterns_pcf}.i)
until $Re < Re_c$ and the flow relaminarizes.
At $Ro=0.2$ we observe oblique patterns
if $250 \lesssim Re \lesssim 600$
with different turbulence activity but without
laminar-like flow regions
(Figure \ref{patterns_pcf}.j).
With increasing $Ro$ the oblique patterns gradually
disappear (figure \ref{patterns_pcf}.k)
and we only see uniform turbulence or regular
vortices when $Re > Re_c$ 
(figure \ref{patterns_pcf}.l).

Figure \ref{trans} shows a survey of the observed flow regimes as a function
of $Re$ and $Ro$ in the DNS of XPPF and XPCF.
\begin{figure}
\begin{center}
\begin{tabular}{cc}
\includegraphics[width=6.5cm]{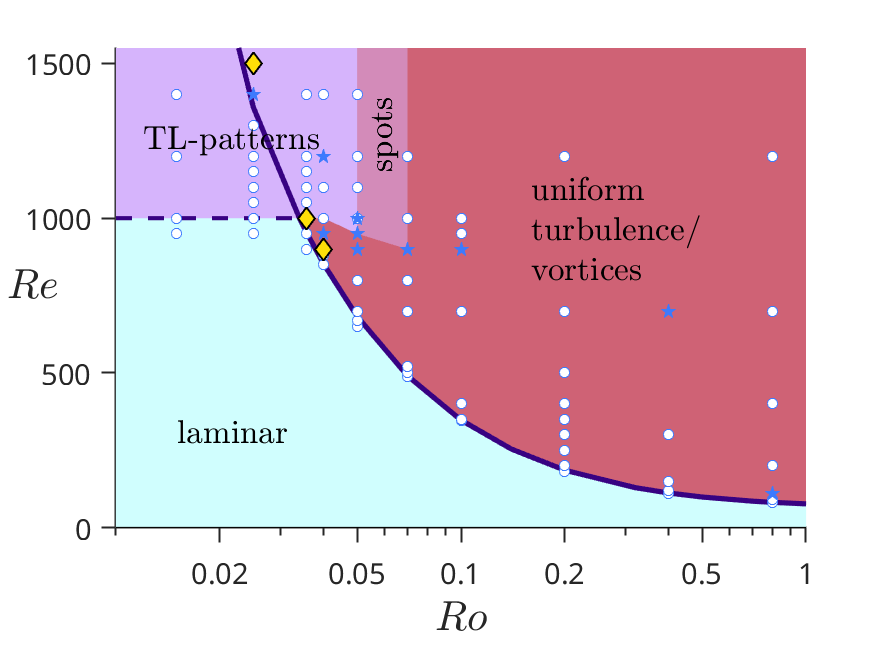}
&
\includegraphics[width=6.8cm]{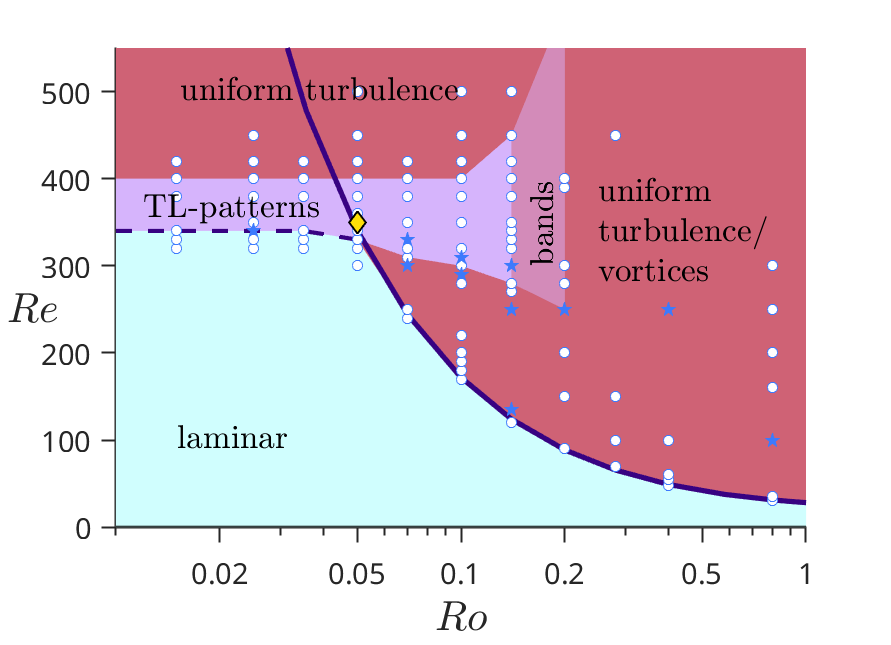}\\
(a) & (b)
\end{tabular}
\end{center}
\caption{Flow regimes as a function of $Ro$ and $Re$ in
(a) XPPF and (b) XPCF. Four flow regimes are
distinguished (eached marked by a different colour): a regime with 
(i) uniform/featureless turbulence or vortices,
(ii) laminar flow,
(iii) turbulent-laminar (TL) patterns,
(iv) spotty structures or spots (XPPF)/band-like structures (XPCF)
but no clear turbulent and laminar flow regions.
Also shown are the neutral stability curve (solid line), 
subcritical threshold $Re_t$ (dashed line),
conditions at which two stable non-laminar flow states coexist (yellow diamonds),
conditions at which DNS were performed (white circles),
and conditions corresponding to the visualizations shown in figures \ref{patterns_ppf} and \ref{patterns_pcf} (blue stars).
\label{trans}}
\end{figure}
We distinguish between four flow regimes: a fully laminar regime; a transitional
regime with local relaminarization or large-scale turbulent-laminar patterns;
a regime with a less clear distinction between turbulent and laminar flow
regions but with large-scale patterns;
and a regime with uniform turbulence or vortical motions.
In XPPF and XPCF, organized and steady vortices appear 
at higher $Ro$ near $Re_c$. As $Re$ increases further,
these vortices gradually become more unsteady and chaotic, ultimately transitioning into a uniformly turbulent flow. As a result, it was not possible to define a sharp transition between the regimes characterized by uniform vortices and uniform turbulence; therefore, these regimes are not treated separately.

Furthermore, we have not observed the variety of vortical structures reported previously for ZPCF \citep{Tsukahara,Suryadi2014}
and Taylor-Couette flow \citep{Andereck}.
Also, no clear qualitative differences were evident in the uniform turbulence regime of XPCF for $Ro\lesssim 0.14$ and only
at higher rotation rates (e.g., $Ro=0.8$),
the effects of rotation on the vortical structures become noticeable (not shown here). Developing a more detailed and refined regime map would require extensive additional simulations and analysis and is therefore beyond the scope of the present study.

At some $Ro$ we observe two coexisting stable regimes in XPPF and XPCF
at a fixed $Re$ somewhat higher than $Re_c$;
a regime with regular vortices nearly aligned with the
streamwise direction but without turbulent motions,
and a regime with transitional or turbulent flow.
This regime with regular vortices appears when the DNS
is initialized with a laminar flow with small noise.
These two coexisting nonlaminar flow regimes are only
observed in a narrow $Ro$ range, see figure \ref{trans}.
This differs from NPPF and NPCF and XPPF and XPCF at low $Ro$
when under subcritical conditions only
a transitional and laminar flow regime are stable.

Figure \ref{trans} shows that
subcritical transition can be triggered at
low $Ro$ in XPPF and XPCF since $Re_t < Re_c$,
while the flow relaminarizes if $Re< Re_t$.
At higher $Ro$, when $Re_t > Re_c$ or when the transitional
regime is absent, we cannot find evidence of
subcritical transition since in all our DNS, XPPF and XPCF
then relaminarize if $Re < Re_c$.
This absence of subcritical transition in XPPF and XPCF at
higher $Ro$ was checked by initializing the DNS in two different ways; 
(i) with a uniformly or transitional turbulent flow at higher $Re$
and subsequently reducing $Re$ in steps until $Re$ was slightly
below $Re_c$, (ii) with a flow with strong disturbances
at $Re$ slightly below $Re_c$. 
In both cases, the flow relaminarized in the DNS.
The crossover from the low-$Ro$ range with subcritical transition
to high-$Ro$ range without subcritical transition
is at $Ro \simeq 0.034$ in XPPF and $Ro\simeq 0.05$ in XPCF.
Observations do not change fundamentally for $Ro>1$, that
is, turbulent motions or vortices only develop if $Re > Re_c$.
In ZPPF and ZPCF, there is likewise no evidence of 
subcritical transition once rotation
has substantially reduced $Re_c$ \citep{Alfredsson,Tsukahara}. 
Moreover, in XPCF,
subcritical transition must vanish entirely in the limit
$Ro\rightarrow\infty$,
since in this limit $Re_c$ and $Re_E$ coincide,
implying that transient growth cannot occur for $Re < Re_c$.
Figure \ref{trans} further shows that
a transitional regime with turbulent-laminar patterns is observed
in XPPF and XPCF, as in NPPF and NPCF,
at low $Ro$ but not at higher $Ro$. 
These patterns develop even though streamwise rotation acts destabilizing
and lowers $Re_c$, while in ZPCF turbulent-laminar patterns
are so far only observed when rotation is cyclonic and stabilizes
the flow \citep{Tsukahara,Brethouwer2012}.
The $Re$ range with turbulent-laminar patterns is fairly constant with $Ro$.
In XPCF we observe a transitional regime at
$340 \lesssim Re \lesssim 400$, and
in XPPF at $Re\gtrsim 1000$ with the upper bound not determined here.
Subcritical transition in XPPF and XPCF is thus only observed
when $Re_c$ is higher than the lower bound for 
turbulent-laminar patterns, that is, when $Re_c \gtrsim 1000$
in XPPF and $Re_c \gtrsim 340$ in XPCF.

Interestingly, we observe a transitional regime with
turbulent-laminar patterns in XPPF and XPCF
in a small range $Ro$ when $Re> Re_c$, unlike
in NPPF and NPCF where this regime only appears
if the flow is subcritical.
This suggests that in this small $Ro$ range,
patterns can emerge by lowering and raising $Re$.
Indeed, in XPPF at $Ro=0.04$ and $0.05$ turbulent laminar patterns emerge
in our DNS starting not only from a turbulent flow at higher $Re$
and subsequently lowering $Re$, but also from a flow with vortices
but without turbulence at lower $Re$ and subsequently increasing $Re$.

In fact, we can observe the formation of turbulent-laminar patterns
at $Re > Re_c$ in
XPPF and XPCF with laminar flow and some noise as initial condition.
To show this, we carry out DNS of XPPF at 
$Re=1200$ and $1500$ and $Ro=0.04$ ($Re_c =852$) and $Ro=0.05$ ($Re_c =683$)
with a computational domain size of $242 \times 2 \times 110$
and resolution $1536 \times 65 \times 1536$ in the 
streamwise, wall-normal and spanwise direction, respectively,
as well as DNS of XPCF at
$Re=350$ and $Ro=0.07$ ($Re_c =244$) and $Ro=0.1$ ($Re_c =172$)
with a computational domain size of $750 \times 2 \times 375$
and resolution $1920 \times 49 \times 1920$ in the 
streamwise, wall-normal and spanwise direction, respectively.
These domains and resolutions are larger than in
our other DNS of XPPF and XPCF to show
the robustness of the observations.
The initial condition is a laminar
base flow with small noise.

Initially, a linear instability occurs in all
six DNS since $Re > Re_c$,
leading to an exponential growth of $u'$ and $v'$, see 
figure \ref{supercritical}(a,b) for XPPF
at $Ro=0.04$ and XPCF at $Ro=0.07$, respectively.
\begin{figure}
\begin{center}
\begin{tabular}{cc}
\includegraphics[width=6.0cm]{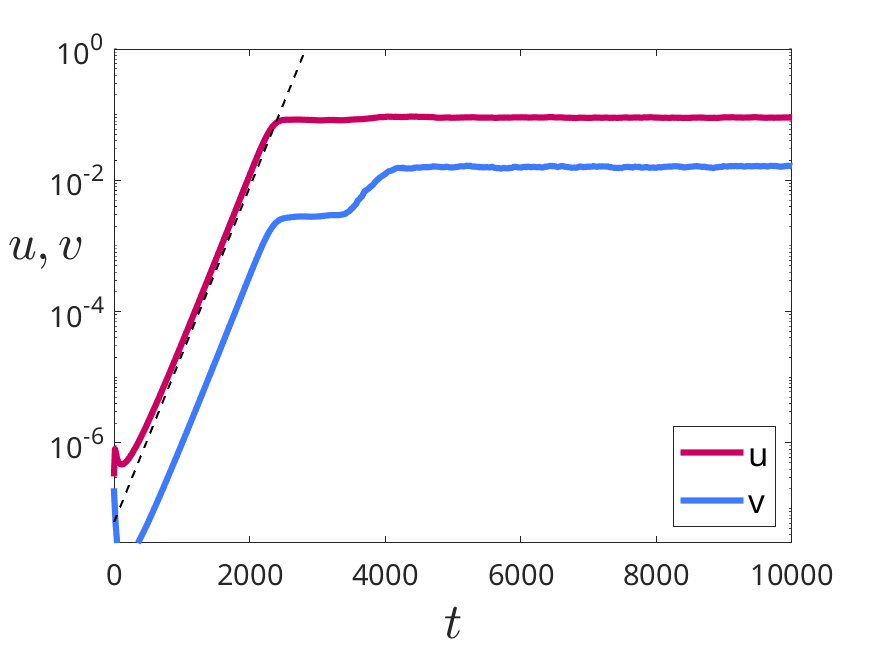}&
\includegraphics[width=6.0cm]{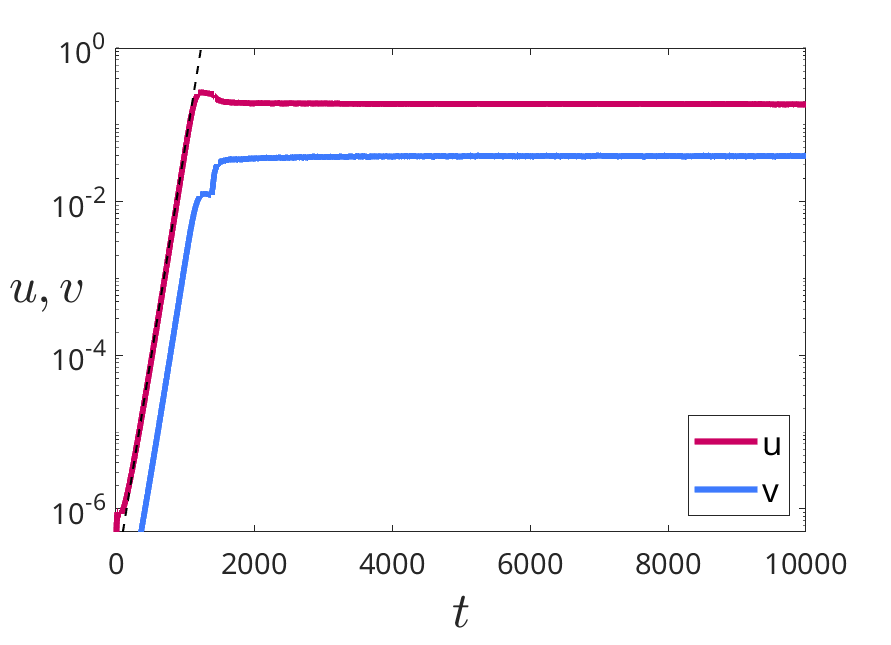}\\
(a) & (b) \\
\includegraphics[width=6.5cm]{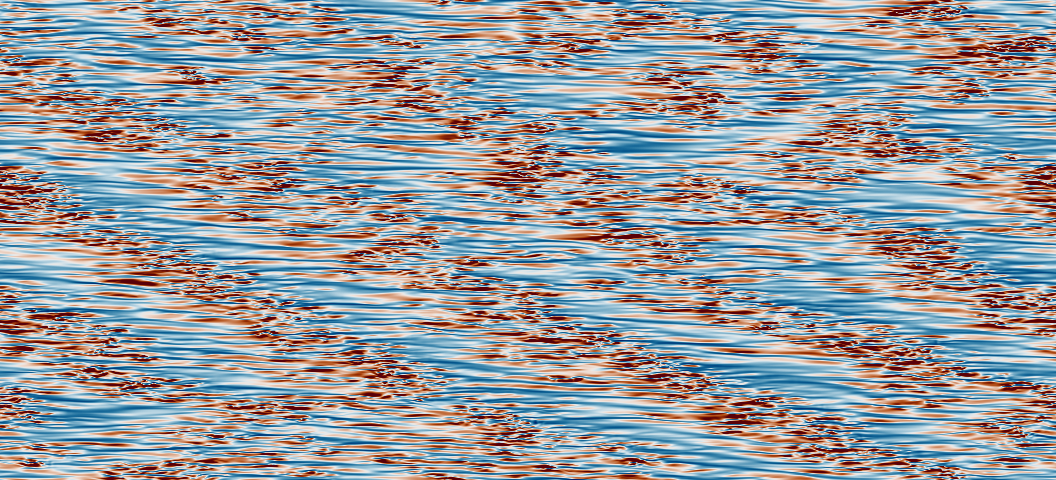}&
\includegraphics[width=6.5cm]{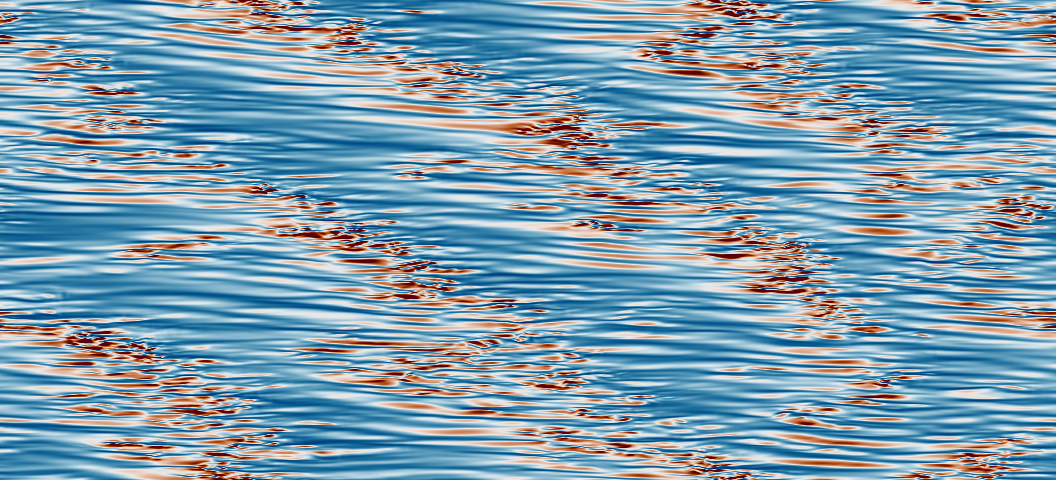}\\
(c) $Ro=0.04$, $Re=1500$ & (d) $Ro=0.05$, $Re=1200$\\
\includegraphics[width=6.5cm]{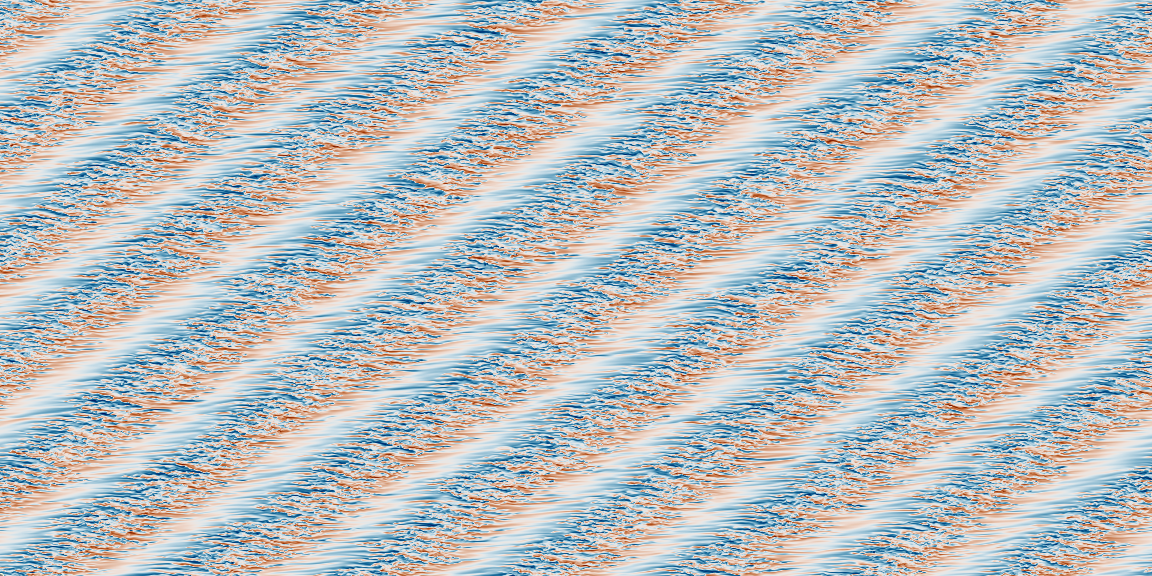}&
\includegraphics[width=6.5cm]{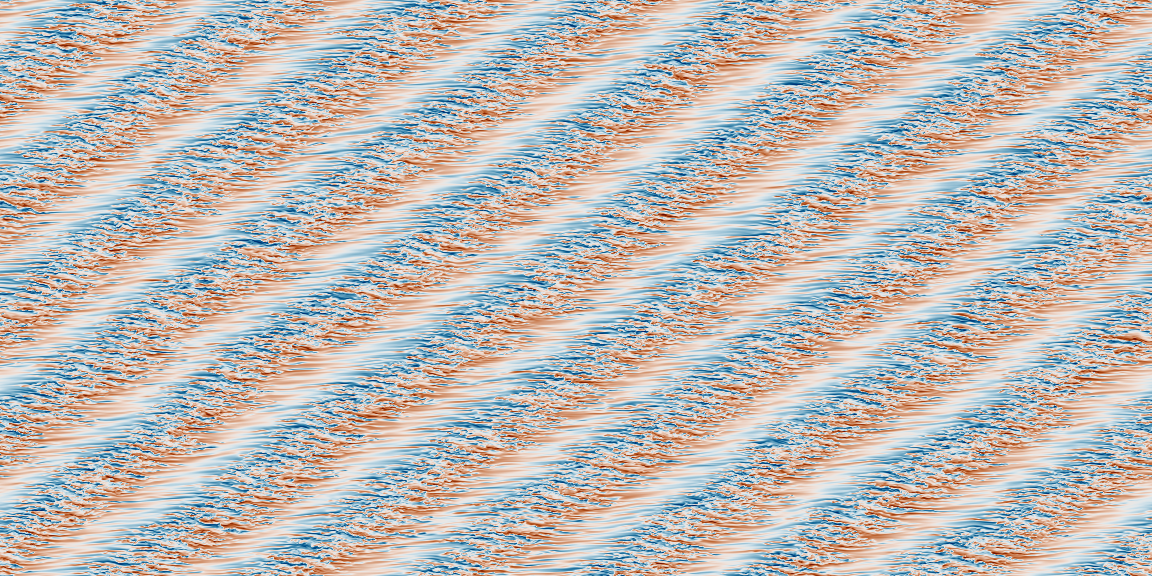}\\
(e) $Ro=0.07$, $Re=350$ & (f) $Ro=0.1$, $Re=350$
\end{tabular}
\end{center}
\caption{Time series of the streamwise (red line)
and wall-normal (pink line) velocity fluctuations in
(a) XPPF at $Ro=0.05$ and $Re=1200$, and 
(b) XPCF at $Ro=0.1$ and $Re=350$.
Visualization of the streamwise velocity field in an $xz$-plane in
at $y=-0.9$ in (c,d) XPPF and
at $y=0$ in (e,f) XPCF. 
\label{supercritical}}
\end{figure}
Here, $u'$ and $v'$ are the streamwise and wall-normal
velocity fluctuations integrated over the whole volume.
After this transient period $u'$ and $v'$ saturate,
and turbulent-laminar patterns appear, which
develop into sustained oblique turbulent-laminar bands in our DNS
of XPPF and XPCF, see figure \ref{supercritical}(c-f),
with similar results for the other two XPPF cases.
These results show
that in a limited $Ro$ range turbulent-laminar patterns
develop under subcritical {\em and}
supercritical conditions in XPPF and XPCF.
By contrast, in DNS at lower $Ro$ with $Re_t < Re< Re_c$,
that is, DNS of XPPF at $Re=1500$, $Ro=0$ ($Re_c=5772$) and
$Re=1200$, $Ro=0.025$ ($Re_c=1359$), and DNS of
XPCF at $Re=350$, $Ro=0$ ($Re_c \rightarrow\infty$) and
$Re=350$, $Ro=0.025$ ($Re_c = 680$) (not shown here),
transition to turbulence
and formation of turbulent-laminar patterns
only occurs when the initial noise levels are finite.

Turbulent–laminar patterns have not yet been observed in ZPCF under supercritical conditions \citep{Tsukahara}, whereas such patterns can develop in ZPPF, though only on the channel side stabilized by rotation \citep{Brethouwer2017}. 
It is possible that the dominant streamwise roll cells, triggered by the strongly destabilizing effect of anti-cyclonic spanwise rotation, inhibit the formation of turbulent–laminar patterns.
In Taylor–Couette flow (TCF), turbulent–laminar bands appear as spiral patterns under subcritical \citep{Meseguer_PhysRevE2009_subcritical,Burin2012} and supercritical conditions \citep{Meseguer_PhysRevE2009_supercritical,Wang_JFM2022}. We note similarities with the XPPF and XPCF cases considered here, in which turbulent–laminar patterns also appear under subcritical and supercritical conditions.

\cite{Berghout2020} and \cite{Wang_PTRoySoc2023} performed DNS of TCF with counter-rotating cylinders in the supercritical regime, analyzing the formation and statistical characteristics of these spiral patterns. Their observed spiral patterns closely resemble those found in subcritical NPCF \citep{Wang_PTRoySoc2023}. However, in TCF, weak vortices persist near the inner cylinder within the laminar-like regions due to the centrifugal instability of the base flow. A similar phenomenon occurs in XPCF under supercritical conditions exhibiting turbulent–laminar patterns; despite significantly weaker fluctuations, streamwise-oriented vortices remain visible within the laminar-like flow regions, see e.g. figure \ref{patterns_pcf}(e,g).

\section{Conclusions}

We carried a linear stability analysis of plane Poiseuille flow (PPF)
and plane Couette flow (PCF) subject to streamwise system rotation.
Linear stability analysis of streamwise rotating PPF has already
been performed by \citet{Masuda}, but we have extended it
and compared it to the PCF case.
Three-dimensional perturbations are considered since
the most unstable modes are inclined to the streamwise direction,
in contrast to spanwise rotating PPF and PCF in which two-dimensional
perturbations with streamwise wavenumber $\alpha=0$ are most unstable
\citep{Lezius,Wall}.

Linear stability analysis of streamwise rotating PCF
shows an asymptotic regime at $Ro \ll 1$
with $Re_c \propto Ro$, and another asymptotic regime
at $Ro \gg 1$ with $Re_c$ approaching a constant value,
as in streamwise rotating PPF \citep{Masuda}.
In both asymptotic regimes, 
the critical spanwise wavenumber $\beta_c$ approaches
a constant value, and the critical vortices become
increasingly streamwise aligned.

The minimum critical Reynolds number $Re_c = 20.66$ of 
streamwise rotating PCF at $Ro \rightarrow \infty$
is equal to the minimum $Re_c$ of spanwise
rotating PCF at $Ro=0.5$ \citep{Lezius}.
Likewise, $Re_c=66.45$ of streamwise rotating
PPF at $Ro \rightarrow \infty$
is equal to the minimum $Re_c$ of spanwise
rotating PPF occurring at $Ro=0.3366$ \citep{Wall}.
These results follow from
the equation for the wall-normal velocity perturbation.
We also show that the linear stability of 
streamwise rotating PCF is related to Rayleigh-B{\'e}nard convection,
like that of spanwise rotating PCF.
In all cases, $\beta_c = 1.558$ and the minimum
$Re_c$ in streamwise and spanwise rotating PCF
at $Ro\rightarrow\infty$ and $Ro=0.5$,
respectively, is related to the critical Raleigh
number $Ra_c$ as $Re_c = \sqrt{Ra_c}/2$.

We carried out DNS of streamwise
rotating PPF and PCF in a range of $Re$ and $Ro$
to investigate flow characteristics at low $Re$
and whether a subcritical transition can occur.
Our DNS show that a subcritical transition can occur
in both flow cases at low $Ro$ but not at higher $Ro$,
since in all
simulations the flow then fully relaminarizes once
$Re < Re_c$. We find that at low $Ro$ the flow
can become transitional and sustained
large-scale turbulent-laminar
patterns can develop at sufficiently low $Re$. These
turbulent-laminar patterns can, especially in
streamwise rotating PCFs, form clear band-like structures.
In a small $Ro$ range, turbulent-laminar patterns
emerge under supercritical conditions when
$Re > Re_c$. We have carried out DNS of
streamwise rotating PPF and PCF to show
that under such conditions
turbulent-laminar patterns can develop from
a growing linear instability when the DNS
are initialized by a laminar flow with small noise.


\backsection[Acknowledgements]{NAISS is acknowledged for providing computational resources in Sweden.}

\backsection[Funding]{This research received financial support from the Swedish Research Council through grant number 2021-03967.}

\backsection[Declaration of interests]{The author reports no conflict of interest.}

\appendix

\section{Comparison LSA and DNS}\label{Ap_A}

To validate the LSA we have performed
DNS of XPCF and XPPF with small
initial perturbations at $Ro=0.05$, $0.8$ and $24$
and $Re$ slightly above $Re_c$. 
Figure \ref{comp} shows the growth of
the root-mean-square of the velocity fluctuations in the
DNS of XPPF and XPCF as well as the growth
rate of the most unstable mode predicted by LSA
at the same $Ro$ and $Re$ as in the DNS.
\begin{figure}
\begin{center}
\begin{tabular}{cc}
\includegraphics[width=6.5cm]{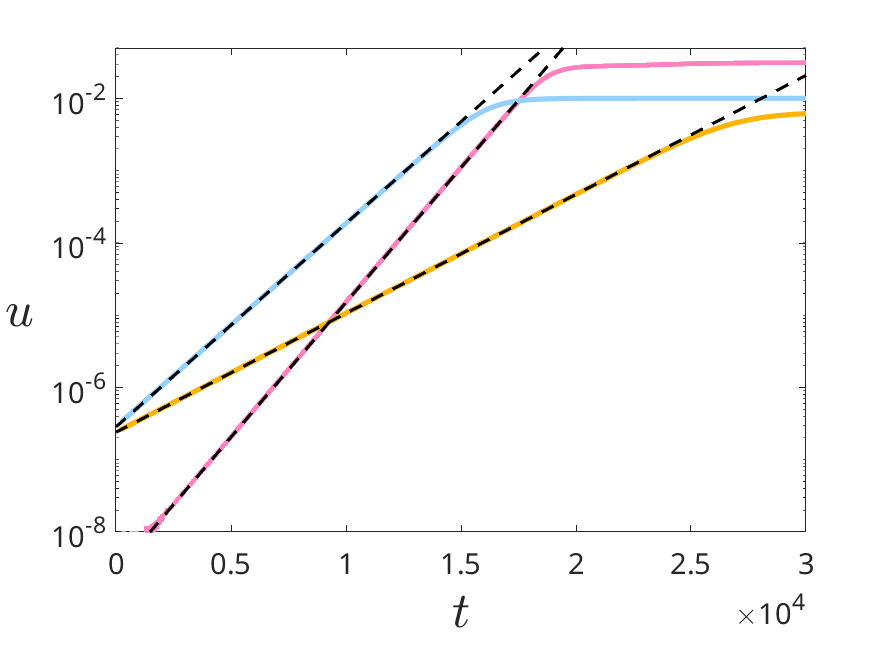}&
\includegraphics[width=6.5cm]{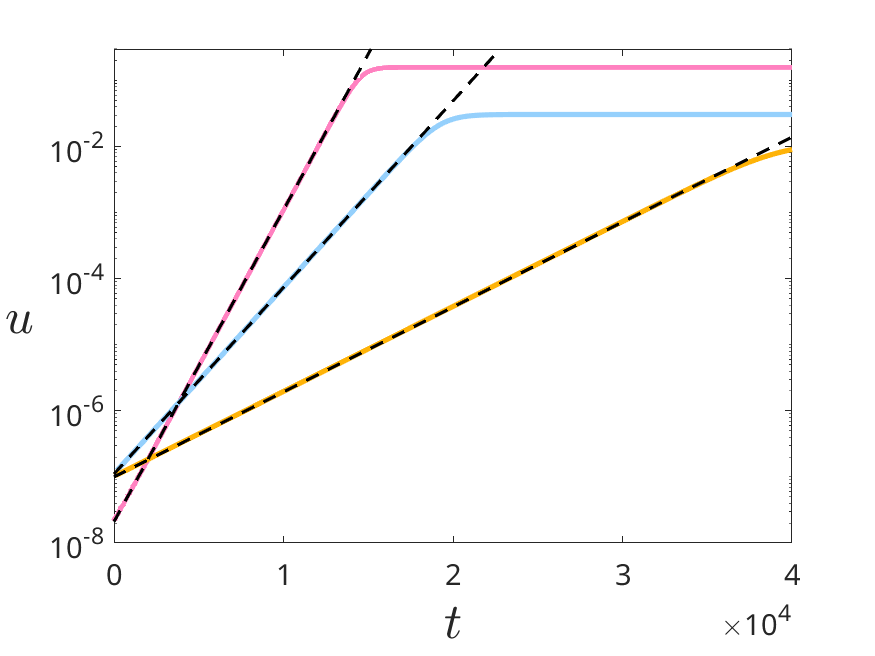}\\
(a) & (b)
\end{tabular}
\end{center}
\caption{Comparison between the growth of the streamwise
velocity fluctuation in DNS of XPPF at
(a) $Ro=0.05$ and $Re=725$ (pink line),
$Ro=0.8$ and $Re=82.2$ (light blue line),
$Ro=24$ and $Re=66.57$ (amber line), 
and DNS of XPCF at
(b) $Ro=0.05$ and $Re=370$ (pink line),
$Ro=0.8$ and $Re=31.35$ (light blue line),
$Ro=24$ and $Re=20.7$ (amber line),
and the growth rate predicted by LSA
at the same $Ro$ and $Re$ (dashed lines).
\label{comp}}
\end{figure}
The computational domain size in all DNS is taken as
$L_x = 4\pi/\alpha$ and $L_z = 8\pi/\beta$,
where $\alpha$ and $\beta$ are the streamwise and
spanwise wavenumber of the most unstable mode,
as predicted by LSA. 
The resolution is $128 \times 97 \times 96$
and $128 \times 65 \times 96$ in the DNS
of XPPF and XPCF, respectively.
Figure \ref{comp} shows that the DNS and LSA
results coincide.

\bibliographystyle{jfm}
\bibliography{ref}
\end{document}